\documentclass[runningheads]{llncs}
\usepackage{graphicx}

\usepackage[latin1]{inputenc}
\usepackage[english]{babel}
\usepackage[bookmarks,bookmarksopen,bookmarksdepth=2]{hyperref}

\usepackage{enumerate}
\usepackage{todonotes}
\usepackage{amsmath}
\usepackage{amsfonts}
\usepackage{amssymb}
\usepackage[ruled,vlined]{algorithm2e}
\SetAlFnt{\small}
\SetAlCapFnt{\large}
\SetAlCapNameFnt{\large}
\usepackage{listings}
\SetAlgoLined
\SetNlSty{texttt}{}{:}
\SetKwProg{Fn}{Fun}{}{}
\SetKwProg{Hdl}{Upon}{}{}
\newcommand*{\N}{\mathbb{N}}

\newcommand{\st}[1]{\ensuremath{\ \mathbf{s.t.} \ #1}}
\newcommand{\from}[1]{\ensuremath{\ \mathbf{from} \ #1}}
\newcommand{\mathsc}[1]{{\ensuremath{\normalfont\textsc{#1}}}}
\newcommand{\majority}[1]{\left\lceil\frac{|#1|+1}{2}\right\rceil}
\newcommand{\id}[1]{\ensuremath{\mathit{#1}}}
\newcommand{\atom}[1]{\ensuremath{\mathsc{#1}}}
\newcommand{\msgatom}[1]{\langle\mathsc{#1}\rangle}
\newcommand{\msg}[2]{\langle\mathsc{#1}\mid#2\rangle}
\newcommand{\send}[2]{\mathbf{send} \ #1 \ \mathbf{to} \ #2}
\newcommand{\trigger}[1]{\mathbf{trigger} \ #1}
\newcommand{\inlineif}[3]{\mathbf{if} \ #1 \ \mathbf{then} \ #2 \ \mathbf{else} \ #3}
\newcommand{\inlineforall}[1]{\ \mathbf{for all} \ #1}
\newcommand{\emptyseq}{\langle\rangle}
\newcommand{\seq}[1]{\langle#1\rangle}
\newcommand{\persist}[1]{\textbf{persistent} \ #1}
\DeclareMathOperator{\append}{++}
\def\HiLi{\leavevmode\rlap{\hbox to \hsize{\color{yellow!50}\leaders\hrule height .8\baselineskip depth .5ex\hfill}}}
\def\ToDo{\leavevmode\rlap{\hbox to \hsize{\color{red!50}\leaders\hrule height .8\baselineskip depth .5ex\hfill}}}

\title{Lecture Notes on\\Leader-based Sequence Paxos}
\subtitle{An Understandable Sequence Consensus Algorithm}
\titlerunning{Leader-based Sequence Paxos}
\author{Seif Haridi\inst{1,2} \and Lars Kroll\inst{2} \and Paris Carbone\inst{1,2}}
\institute{KTH Royal Institute of Technology, Stockholm, Sweden\\
\and
RISE Research Institutes of Sweden, Stockholm, Sweden\\
\email{\{haridi, lkroll, parisc\}@kth.se}
}

\begin{document}

\maketitle

\begin{abstract}
Agreement among a set of processes and in the presence of partial failures is one of the fundamental problems of distributed systems. In the most general case, many decisions must be agreed upon over the lifetime of a system with dynamically changing membership. Such a sequence of decisions represents a distributed log, and can form the underlying abstraction for driving a replicated state machine. While this abstraction is at the core of many systems with strong consistency requirements, algorithms that achieve such sequence consensus are often poorly understood by developers and have presented a significant challenge to many students of distributed systems. In these lecture notes we present a complete and practical Paxos-based algorithm for reconfigurable sequence consensus in the fail-recovery model, and a clear path of simple step-by-step transformations to it from the basic Paxos algorithm.
\end{abstract}

\section{Introduction}
\label{sec:intro}
Agreement among a set of processes one of the fundamental problems of distributed systems. The challenges arise mostly from the possibility of \emph{partial failures} that differentiate distributed programming from parallel programming. Under such conditions, not only must algorithms preserve safety guarantees despite failing nodes, but over the lifetime of a real system, the involved nodes must change to compensate for failures. Thus, in the most general case, not one, but many decisions must be agreed upon by a dynamically changing set of processes. A sequence of such agreed-upon decisions forms a \emph{distributed log}, and often forms the basis upon which we build abstractions such as a \emph{replicated state machine} (RSM). RSMs can be used as replicated databases~\cite{kroll2013load}, lock services~\cite{burrows2006chubby}, or configuration management services~\cite{hunt2010zookeeper}, for example.

While this replicated log abstraction is at the core of many such systems with strong consistency requirements, the actual algorithms that achieve what we call \emph{sequence consensus} are often poorly understood by developers, and have presented a significant challenge to many students (and teachers) of distributed systems over the years.

In these notes we will incrementally describe an algorithm for reconfigurable sequence consensus in the fail-recovery model~\cite{cachin2011introduction}. Our algorithm is based on the well-known Paxos algorithm by Leslie Lamport~\cite{lamport1998part}, which we will present as a starting point in section \ref{sec:paxos}. In section \ref{sec:sequencepaxos} we will describe the sequence consensus abstraction and a first simple algorithm to implement it. To improve our implementation, we will first take a detour into leader election in section \ref{sec:ble}, before using section \ref{sec:leaderpaxos} to reduce the communication cost and the memory footprint of our algorithm. Once we have a working and efficient algorithm in the \emph{fail-stop} model, we will extend our implementation to function in the fail-recovery model in section \ref{sec:recovery}, and particularly describe how to deal with (TCP) link session loss. In section \ref{sec:reconf} we extend our algorithm to allow the introduction of new processes, by describing how to move from one \emph{system configuration} to the next. As the algorithm uses some state with unbounded growth at this point, we discuss garbage collection mechanisms in section \ref{sec:gc}, before discussing literature and concluding in sections \ref{sec:lit} and \ref{sec:conc} respectively.

\section{Paxos}
\label{sec:paxos}
Given a set of processes in a partially-synchronous system model, i.e. an asynchronous system with stable periods of ``sufficient'' length, we wish a single value $v$ to be agreed upon. That is, all processes should ``decide'' on the same value, such that the following properties hold:
\begin{description}
\item[UC1 (Validity)] Only proposed values may be decided.
\item[UC2 (Uniform Agreement)] No two processes decide different values.
\item[UC3 (Integrity)] Each process can decide a value at most once.
\item[UC4 (Termination)] Every correct process eventually decides a value.
\end{description}
As part of the model we are given a channel abstraction that allows message duplication, losses, and out-of-order delivery. On top of the $\diamond P$ failure detection abstraction, we use an eventual leader election abstraction $\Omega$. The Paxos algorithm provides the \textbf{UC} properties by using $\Omega$'s leader to impose a value to be decided. The algorithm guarantees safety during unstable periods and $\Omega$ provides liveness during a stable period. In Paxos each process plays one or more, quite often all, of the following roles: 
\begin{description}
\item[Proposer] Wants a particular proposed value to be decided.
\item[Acceptor] Acknowledges acceptance of proposed values.
\item[Learner] Decides based on acceptance of values.
\end{description}
In the typical majority quorum setup, a \emph{proposer} tries to get a majority of \emph{acceptors} to accept its proposal $v$. If a proposal has a majority of acceptors, then it is called \emph{chosen} and the \emph{learners} will decide it, once they discover that this is the case. Other quorum variants than majority have also been proposed for use with Paxos, and can improve performance at the cost of resilience. For brevity we will only discuss majority quorums in this article.\\

The process described at a high level above, is split into two phases. Multiple instance of each phase can run concurrently during any execution. Many of these instances may abort, until eventually a single $v$ is decided.

\paragraph{Prepare Phase}
A \emph{proposer} starts by picking a unique sequence number $n$ and sending a message $\msg{Prepare}{n}$ to all acceptors. Upon receiving such a message an \emph{acceptor} will either promise to not accept any proposal with a sequence number $n'<n$ or ignore/refuse the prepare, if it has already promised the same to someone with a higher $n$. If it promises, it will reply with a message $\msg{Promise}{n', v'}$, where $v'$ is the highest numbered proposal it has accepted so far (if any) and $n'<n$ its proposal number. The \emph{proposer} collects all the promises it receives until it has a majority.
\paragraph{Accept Phase}
Once a \emph{proposer} has collected a majority $S$ of promises, it picks the highest numbered value $v$ in $S$ -- or whatever value it wishes to propose, if there are no values in $S$ -- and sends a message $\msg{Accept}{n,v}$ to all acceptors. When an \emph{acceptor} receives such a message, it replies with a simple $\msgatom{Ack}{}$ message, unless it has issued a higher numbered promise in the meantime, in which case it will reject with a $\msgatom{Nack}{}$. If the \emph{learners} are separate processes from the \emph{proposer} the $\msgatom{Ack}{}$ messages will need to be broadcast, otherwise they can just be sent to the relevant \emph{proposer} acting in both roles. In either case, once a majority of acks is collected $v$ can be decided, usually via broadcasting $\msg{Decide}{v}$. If nacks make a majority impossible, the procedure must be aborted and started over.\\

During unstable periods the algorithm is guaranteed to satisfy its safety conditions (\textbf{UC1}-\textbf{UC3}), while termination is guaranteed only if the stable period is long enough for a solo proposer to perform the prepare and accept phases with no contention.

\subsubsection{Fail-recovery}
In order to work in the \emph{fail-recovery} model, acceptors have to commit some of their state to stable storage and restore it during recovery. Concretely acceptors need to store the \emph{highest proposal} $(n', v')$ they accepted and the \emph{highest sequence number} $n$ they promised.

\subsubsection{Optimisations} All $\msgatom{Nack}{}$ messages above are technically optimisations, as they can be replaced with timeouts on the proposer side.\\
In addition to the necessary rejection of any accepts with $n<m$ where the acceptor previously promised $m$, there are a number of optimisations that cause earlier aborts and thus waste less time on attempts already doomed to fail.
\begin{enumerate}[a)]
\item Reject $\msg{Prepare}{n}$ if answered $\msg{Prepare}{m}$ with $m > n$.
\item Reject $\msg{Accept}{n, v}$ if answered $\msg{Accept}{m, u}$ with $m > n$.
\item Reject $\msg{Prepare}{n}$ if answered $\msg{Accept}{m, u}$ with $m > n$.
\end{enumerate}
Additionally, the algorithm should ignore any old messages for proposals that already got a majority.\\
Since a value ``chosen'' will always be decided in any higher round, a proposer can also skip the \emph{accept phase} if a majority of acceptors return the same $v$ in the \emph{prepare phase}.\\

\subsubsection{Algorithm} Algorithm \ref{alg:paxosprop} shows the \emph{fail-stop} version of Paxos described above with some of the optimisations. For simplicity of presentation the algorithm uses Perfect Links and no $\Omega$ and there are thus cases where it would never terminate, but end up in a race condition. Augmentation for Fair-loss Links and $\Omega$ is straight forward, if somewhat tedious, though. With Fair-loss Links messages can be lost and duplicated, thus augmentation requires resending, duplicate filtering, and also incrementing timeouts while waiting for majorities to avoid getting stuck. For $\Omega$ the required change is simply to only allow proposals while being leader. As $\Omega$ guarantees to eventually have a single leader, the race condition from before is thus avoided.

\begin{algorithm}[htp]
\caption{Abortable Paxos -- Proposer}
\label{alg:paxosprop}
\textbf{Implements:} Uniform Consensus\\
\textbf{Requires:} Perfect Link \\
\textbf{Algorithm:} \\
\nl$A$ \tcc*{set of acceptors}
\nl$L$ \tcc*{set of learners}
\nl$s$ $\gets 0$ \tcc*{local sequence number}
\nl$(n_p, v_p) \gets (\bot, \bot)$ \tcc*{unique round number and value}
\nl$\id{promises}$ $\gets \emptyset$\;
\nl$\id{acks}$ $\gets 0$\;
\hrulefill \\
\nl\Hdl{$\msg{Propose}{v}$}
{
\nl$s \gets s+1$\\ 
\nl$n_p \gets \atom{unique}(s)$ \tcc*{use pid to make $n$ globally unique}
\nl$v_p \gets v$\\
\nl$\id{promises}$ $\gets \emptyset$\;
\nl$\id{acks}$ $\gets 0$\;
\ForEach {$a \in A$}{
\nl$\send{\msg{Prepare}{n_p}}{a}$\;
}
}
\nl\Hdl{$\msg{Promise}{n, n', v'} \from{a} \st{n=n_p}$}
{
\nl$\id{promises} \gets \id{promises} \cup \{(a, n', v')\}$\tcc*{add $a$ for acceptor disambiguation}
\nl \If{$|\id{promises}| = \majority{A}$}{
\nl$v \gets \atom{maxValue}(\id{promises})$\tcc*{value with the largest $n$}
\nl$v_p \gets \inlineif{v \neq \bot}{v}{v_p}$\label{algl:paxosadopt}\tcc*{adopt $v$ if present}
\ForEach {$a \in A$}{
\nl$\send{\msg{Accept}{n_p, v_p}}{a}$\;
}
}
}
\nl\Hdl{$\msg{Ack}{n} \from{a} \st{n=n_p}$}
{
\nl$\id{acks} \gets \id{acks} + 1$\;
\nl \If{$\id{acks} = \majority{A}$}{
\ForEach {$l \in L$}{
\nl$\send{\msg{Decide}{v_p}}{l}$\;
}
}
}
\nl\Hdl{$\msg{Nack}{n} \from{a} \st{n=n_p}$}
{
\nl$\atom{abort}()$ \tcc*{Goto $\msg{Propose}{v}$ and pick a new $n_p$ immediately to avoid old messages being handled}
}
\end{algorithm}
\addtocounter{algocf}{-1}
\begin{algorithm}[htp]
\caption{Abortable Paxos -- Acceptor}
\label{alg:paxosacc}
\textbf{Implements:} Uniform Consensus\\
\textbf{Requires:} Perfect Link \\
\textbf{Algorithm:} \\
\nl$n_{prom}\gets 0$ \tcc*{promise not to accept in lower rounds}
\nl$(n_a, v_a) \gets (\bot, \bot)$ \tcc*{sequence number and value accepted}
\hrulefill \\
\nl\Hdl{$\msg{Prepare}{n} \from{p}$}
{
\nl \eIf{$n_{prom} < n$}{
\nl$n_{prom} \gets n$\;
\nl$\send{\msg{Promise}{n, n_a, v_a}}{p}$\;
}{
\nl$\send{\msg{Nack}{n}}{p}$\tcc*{optimisation only}
}
}
\nl \Hdl{$\msg{Accept}{n, v} \from{p}$}
{
\nl \eIf{$n_{prom} \leq n$}{
\nl$n_{prom} \gets n$\;
\nl$(n_a, v_a) \gets (n, v)$\;
\nl$\send{\msg{Accepted}{n}}{p}$\;
}{
\nl$\send{\msg{Nack}{n}}{p}$\tcc*{optimisation only}
}
}
\end{algorithm}
\addtocounter{algocf}{-1}
\begin{algorithm}[htp]
\caption{Abortable Paxos -- Learner}
\label{alg:paxoslearn}
\textbf{Implements:} Uniform Consensus\\
\textbf{Requires:} Perfect Link \\
\textbf{Algorithm:} \\
\nl$v_d$ $\gets \bot$ \tcc*{decided value}
\hrulefill \\
\nl\Hdl{$\msg{Decide}{v}$}
{
\nl \If{$v_d = \bot$}{
\nl$v_d \gets v$\;
\nl$\trigger{\msg{Decide}{v_d}}$\;
}
}
\end{algorithm}

\clearpage

\section{Sequence Paxos}
\label{sec:sequencepaxos}
If our goal is to build a replicated state machine (RSM) based on Paxos, deciding a single value will not suffice. Instead we want to agree on a \emph{sequence of values} that is fed into a deterministic automaton on each replica, such that replicas that have seen the same sequence will have the same state (assuming they started from the same initial state). Such a sequence of values can be seen as a \emph{replicated log} of state machine \emph{commands}.

\subsubsection{Na\"{i}ve Approach} One way to extend Paxos for this scenario is to augment the proposals with instance numbers and run a single value Paxos for each instance in sequential rounds. Thus, at round $i$ each process starts a new instance of Paxos. If it has commands it wants to propose (in a set $\id{proCmds}$) and it has not proposed (variable $\id{proposed} = \atom{false}$) in the current round already, it proposes some command $C\in \id{proCmds}$ as $\msg{Propose}{C, p, i}$ (where $p$ is the process id of the client that sent $C$) and sets $\id{proposed} \gets \atom{true}$. Once it sees a $\msg{Decide}{C', p', i}$, it removes $(C', p')$ from $\id{proCmds}$ and appends $(C', p', i)$ to $\id{log}$. It then executes the command on the state machine $(s_i, res_i) = C(s_{i-1})$, and returns $res_i$ to $p'$. At this point the round ends, it resets $\id{proposed} \gets \atom{false}$ and moves to the next round $i+1$.

The issue with this approach is, that it is completely sequential, working on one round after the other and taking (at least) 4 communication steps (2 round-trips) for each round. Trying to improve performance by pipelining is not straight-forward, as duplicate command entries or log holes must be avoided. However, the obvious optimisation of preparing multiple instance ahead of time and only running the \emph{accept phase} sequentially, since $v$ is not needed in the \emph{prepare phase}, halves the required number of communication steps on the ``hot path'' (and with some batching also reduces it overall)~\cite{lamport2001paxos}.

\subsection{Sequence Consensus}
\label{sec:seqcon}
In order to match our abstraction better with the requirements of a replicated log, a change in interface and desired properties is needed. We will still propose a single command $C$, but we now decide on a sequence of commands $CS$. The original \emph{Uniform Consensus} properties UC1-4 are altered as shown below:

\begin{description}
\item[SC1 (Validity)] If process $p$ decides $CS$ then $CS$ is a sequence of proposed commands (without duplicates\footnote{It is also possible to allow duplicates in the log and filter them out at the state-machine level instead. This simplifies the implementation.}).
\item[SC2 (Uniform Agreement)] If process $p$ decides $CS$ and process $q$ decides $CS'$ then one is a prefix of the other.
\item[SC3 (Integrity)] If process $p$ decides $CS$ and later decides $CS'$ then $CS$ is a strict prefix of $CS'$.
\item[SC4 (Termination)] If a command $C$ is proposed infinitely often by a correct process, then eventually every correct process decides a sequence containing $C$. If duplication is allow, then the decided sequence will contain $C$ infinitely often.
\end{description}

\subsection{Initial Sequence Paxos Implementation}
\label{ssec:seqpax1}
To make it easy to see that the algorithm is correct, we will start with a very simple and inefficient variant of single value Paxos to implement \emph{Sequence Consensus}, and then later add optimisation transformations to it step by step, preserving correctness with each change. We start with the basic \emph{Paxos} presented in algorithm \ref{alg:paxosprop} and make the following changes: All values are now sequences and the empty value $\bot$ becomes the empty sequence $\emptyseq$. After adopting the sequence (value) with the highest proposal number (alg. \ref{alg:paxosprop} l. \ref{algl:paxosadopt}), the sequence is extended by one or more new commands instead of replacing the value. Instead of deciding only if there has been no previous decision, learners will now decide whenever the received sequence is longer than the previously decided one. In order to abstract over the \textbf{SC1} variants with or without duplication we use the \emph{append} operator $\oplus$ with the following two definition variants:
\begin{description}
\item[No Duplicates] \begin{align*}
\seq{C_1, \ldots, C_n} \oplus C &\overset{\id{def}}{=} \left\{\begin{array}{ll}
        \seq{C_1, \ldots, C_n}, & \text{if } C \text{ is equal to some } C_i\\
         \seq{C_1, \ldots, C_n, C}, & \text{otherwise}
        \end{array}\right.
\end{align*}
\item[Duplicates Allowed] \begin{align*}
\seq{C_1, \ldots, C_n} \oplus C &\overset{\id{def}}{=} \seq{C_1, \ldots, C_n, C}
\end{align*}
\end{description}
Algorithm \ref{alg:seqpaxos1prop} shows the described implementation. 
\subsubsection{Correctness} The only changes that have been made affect how values are treated, while the round numbers have been left untouched. The same mechanism ensuring that chosen values are never replaced, now ensures that chosen sequences are always extended and thus no sub-sequences disappear. It is easy to see that this algorithm is as correct as algorithm \ref{alg:paxosprop}.
\subsubsection{Performance} As far as efficiency is concerned, however, we have gained nothing, so far, over the na\"{i}ve multi-instance Paxos described above. In fact, algorithm \ref{alg:seqpaxos1prop} is actually worse, as it sends whole sequences of commands in every step, so its performance actually degrades with the growth of the log.\\

Before we get to a more efficient leader-based implementation where all roles run in each process, we must first make a short excursion into leader election.

\begin{algorithm}[htp]
\caption{Initial Sequence Paxos -- Proposer}
\label{alg:seqpaxos1prop}
\textbf{Implements:} Sequence Consensus\\
\textbf{Requires:} Perfect Link \\
\textbf{Algorithm:} \\
\nl$A$ \tcc*{set of acceptors}
\nl$L$ \tcc*{set of learners}
\nl$s$ $\gets 0$ \tcc*{local sequence number}
\nl$(n_p, v_p) \gets (\bot, \emptyseq)$ \tcc*{unique round number and sequence}
\nl$C_p \gets \bot$ \tcc*{command we are currently trying to append}
\nl$\id{promises}$ $\gets \emptyset$\;
\nl$\id{acks}$ $\gets 0$\;
\hrulefill \\
\nl\Hdl{$\msg{Propose}{C}$}
{
\nl$s \gets s+1$\\ 
\nl$n_p \gets \atom{unique}(s)$ \tcc*{use pid to make $n$ globally unique}
\nl$C_p \gets C$\\
\nl$\id{promises}$ $\gets \emptyset$\;
\nl$\id{acks}$ $\gets 0$\;
\ForEach {$a \in A$}{
\nl$\send{\msg{Prepare}{n_p}}{a}$\;
}
}
\nl\Hdl{$\msg{Promise}{n, n', v'} \from{a} \st{n=n_p}$}
{
\nl$\id{promises} \gets \id{promises} \cup \{(a, n', v')\}$\tcc*{add $a$ for acceptor disambiguation}
\nl \If{$|\id{promises}| = \majority{A}$}{
\nl$v \gets \atom{maxValue}(\id{promises})$\tcc*{sequence with the largest $n$}
\nl$v_p \gets v \oplus C$\tcc*{adopt $v$ and append}
\ForEach {$a \in A$}{
\nl$\send{\msg{Accept}{n_p, v_p}}{a}$\;
}
}
}
\nl\Hdl{$\msg{Ack}{n} \from{a} \st{n=n_p}$}
{
\nl$\id{acks} \gets \id{acks} + 1$\;
\nl \If{$\id{acks} = \majority{A}$}{
\nl $C_p \gets \bot$\;
\ForEach {$l \in L$}{
\nl$\send{\msg{Decide}{v_p}}{l}$\;
}
}
}
\nl\Hdl{$\msg{Nack}{n} \from{a} \st{n=n_p}$}
{
\nl$\atom{abort}()$ \tcc*{Goto $\msg{Propose}{C}$ and pick a new $n_p$ immediately to avoid old messages being handled}
}
\end{algorithm}
\addtocounter{algocf}{-1}
\begin{algorithm}[htp]
\caption{Initial Sequence Paxos -- Acceptor}
\label{alg:seqpaxos1acc}
\textbf{Implements:} Sequence Consensus\\
\textbf{Requires:} Perfect Link \\
\textbf{Algorithm:} \\
\nl$n_{prom}\gets 0$ \tcc*{promise not to accept in lower rounds}
\nl$(n_a, v_a) \gets (\bot, \emptyseq)$ \tcc*{round number and sequence accepted}
\hrulefill \\
\nl\Hdl{$\msg{Prepare}{n} \from{p}$}
{
\nl \eIf{$n_{prom} < n$}{
\nl$n_{prom} \gets n$\;
\nl$\send{\msg{Promise}{n, n_a, v_a}}{p}$\;
}{
\nl$\send{\msg{Nack}{n}}{p}$\tcc*{optimisation only}
}
}
\nl \Hdl{$\msg{Accept}{n, v} \from{p}$}
{
\nl \eIf{$n_{prom} \leq n$}{
\nl$n_{prom} \gets n$
\nl$(n_a, v_a) \gets (n, v)$\;
\nl$\send{\msg{Accepted}{n}}{p}$\;
}{
\nl$\send{\msg{Nack}{n}}{p}$ \tcc*{optimisation only}
}
}
\end{algorithm}
\addtocounter{algocf}{-1}
\begin{algorithm}[htp]
\caption{Initial Sequence Paxos -- Learner}
\label{alg:seqpaxos1learn}
\textbf{Implements:} Sequence Consensus\\
\textbf{Requires:} Perfect Link \\
\textbf{Algorithm:} \\
\nl$v_d$ $\gets \emptyseq$ \tcc*{decided sequence}
\hrulefill \\
\nl\Hdl{$\msg{Decide}{v}$}
{
\nl \If{$|v_d| < |v|$}{
\nl$v_d \gets v$\;
\nl$\trigger{\msg{Decide}{v_d}}$\;
}
}
\end{algorithm}

\clearpage

\section{Ballot Leader Election}
\label{sec:ble}
As described at the end of section \ref{sec:paxos}, the algorithms presented so far may actually never terminate under the given assumptions without the use of an $\Omega$ leader election abstraction. In this section we want to integrate $\Omega$ into the Sequence Paxos algorithm \ref{alg:seqpaxos1prop}, but at the same time we want to outsource the generation of ballot numbers to the leader election, in order to make the replicated log part of the algorithm easier to follow. Thus, the idea is to elect a leader together with a ballot number that is globally unique and locally monotonically increasing. This leader-ballot pair will then be used by the Sequence Paxos algorithm to start a prepare phase. The new abstraction is called \emph{Ballot Leader Election} (\emph{BLE}) and has a single event $\msg{Leader}{p, n}$ where $p$ is the process that is now leader and $n$ its ballot number. \emph{BLE} implementations must fulfil the following properties, which are an extension of $\Omega$'s properties:
\begin{description}
\item[BLE1 (Completeness)] Eventually, every correct process elects some correct process, if a majority of processes is correct.
\item[BLE2 (Eventual Agreement)] Eventually, no two correct processes elect different correct processes.
\item[BLE3 (Monotonic Unique Ballots)] If a process $L$ with ballot $n$ is elected as leader by a process $p$, then all previously elected leaders by $p$ have ballot numbers $m$ with $m<n$, and the pair $(L, n)$ is globally unique.
\end{description}
In the implementation we will allow a process $p$ to ``inaccurately'' drop a correct leader, as long as the new leader has a higher ballot number. We will also require that a process is elected as a leader only if a majority of processes are correct and alive. As this is anyway required for Sequence Paxos, it does not constitute a limitation in any noticeable way.\\
We will start by assuming a fail-noisy model, that is processes fail by crashing, partially-synchronous system model, and perfect links channel abstraction. However, the final algorithm will turn out to actually work fine in a slightly weaker model that allows message loss and crash-recovery.

\subsection{Gossip Leader Election}
The basic idea for the algorithm is as follows: Each process $p$ has its own unique ballot $n$ formed from a sequence number $s$ and its process id, such that $n=(s, \id{pid}_p)$. This pair is trivially unique, as $\id{pid}_p$ is unique. For an implementation this can be folded into a single (potentially long) number, by taking the size of the process id set $\Pi$ and multiplying it with $s$, such that $n=s\cdot|\Pi| + \id{pid}_p$. If $\Pi$ is not known a-priori, any number guaranteed to be larger than $|\Pi|$ can be substituted, for example \lstinline!Int.MAX!. Given this, each process gossips the its ballot number along with the usual failure-detection heartbeats (with a repeating delay, adjustable by a constant $\Delta$) to all other processes. Eventually, each correct process will elect the process with the highest rank (max ballot), given good network conditions (\textbf{BLE2}). However, a process will only trust a leader, if the leader's ballot is among the collected max ballots from a majority of processes. If a process does not find its current leader's ballot in that set, it will increase its own sequence number $s$, recalculate its ballot $n$ as above and wait until a new ballot gets a majority. This satisfies \textbf{BLE3} and also \textbf{BLE1} assuming a sufficiently long stable period as per the partially-synchronous model. Algorithm \ref{alg:gossiple} shows the pseudocode for the described implementation, with the ballot generation part hidden behind the $\atom{increment}(\id{ballot})$ function.

\begin{algorithm}[htp]
\caption{Gossip Leader Election}
\label{alg:gossiple}
\textbf{Implements:} Ballot Leader Election\\
\textbf{Requires:} Perfect Link \\
\textbf{Algorithm:} \\
\nl$\Pi$\tcc*{Process set}
\nl$\id{round} \gets 0$ \tcc*{round number}
\nl$\id{ballots} \gets \emptyset$\;
\nl$n \gets (0, \id{pid})$ \tcc*{ballot number}
\nl$L \gets \bot$ \tcc*{leader}
\nl$n_{max} \gets n$ \tcc*{largest ballot number seen}
\nl$d \gets \Delta$ \tcc*{heartbeat delay}
\nl$\atom{startTimer}(d)$\tcc*{schedule a timeout event in $d$ timeunits}
\hrulefill \\
\nl\Fn{$\atom{checkLeader}()$}{
\nl$\id{top} = (\id{topProcess}, \id{topN}) \gets \atom{maxByBallot}(\id{ballots} \cup \{(\id{self}, n)\})$\;
\eIf{$\id{topN} < n_{max}$}{
\While{$n \leq n_{max}$}{
\nl$n \gets \atom{increment}(n)$\;
}
\nl$L \gets \bot$\;
}{
\If{$top \neq L$}{
\nl$n_{max} \gets \id{topN}$\;
\nl$L = top$\;
\nl$\trigger{\msg{Leader}{\id{topProcess}, \id{topN}}}$\;
}
}
}

\nl\Hdl{$\msgatom{Timeout}$}
{
\If{$|\id{ballots}| + 1 > \majority{\Pi}$}{
\nl$\atom{checkLeader}()$\;
}
\nl$\id{ballots} \gets \emptyset$\;
\nl$\id{round} \gets \id{round} + 1$\;
\ForEach {$p \in \Pi \st{p \neq \id{self}}$}{
\nl$\send{\msg{HeartbeatRequest}{\id{round}, n_{max}}}{p}$\;
}
\nl$\atom{startTimer}(d)$\;
}

\nl\Hdl{$\msg{HeartbeatRequest}{r, b_{max}} \from{p}$}
{
\If{$b_{max} > n_{max}$}{
\nl$n_{max} \gets b_{max}$\;
}
\nl$\send{\msg{HeartbeatReply}{r, n}}{p}$\;
}

\nl\Hdl{$\msg{HeartbeatReply}{r, b} \from{p}$}
{
\eIf{$r = \id{round}$}{
\nl$\id{ballots} \gets \id{ballots} \cup \{(p, b)\}$\;
}{
\nl${d \gets d + \Delta}$ \tcc*{Increase delay to make sure all replies from the current round are received within the time window.}
}
}
\end{algorithm}
\clearpage

\section{Leader-based Sequence Paxos}
\label{sec:leaderpaxos}
In this section we will make step-by-step transformations to improve the performance of Sequence Paxos.

\subsection{BLE and Uniform Processes}
Now that we have a leader election abstraction that also provides ballot numbers, we will adapt the Sequence Paxos algorithm from section \ref{ssec:seqpax1} to use \emph{BLE}. At the same time we will make an additional assumption, that all roles are available for every process, i.e. each process is a proposer, acceptor, and learner. This a common way of running a log replication service in practice, and it will allow each process (a replica) to share the acceptor and learner state information. At each process $p$ we also introduce a $\id{state}$ variable, that tracks both $p$'s current role in the algorithm, $\atom{leader}$ or $\atom{follower}$, and the phase it is currently in, $\atom{prepare}$ or $\atom{accept}$ (or none $\bot$). Every process starts as a $\atom{follower}$ and can move to a $\atom{leader}$ state by being elected by \emph{BLE}, in which case it will stay until overrun by another leader, at which point it will revert to acting as a follower.\\
As we outsourced the leader election part to \emph{BLE} now, we will also introduce some optimisations that assume leaders are long-lived. Particularly, we want to pipeline $\atom{Accept}$ messages while in the \emph{accept phase}. That is, a leader that completed the \emph{prepare phase} will only send $\atom{Accept}$ for every proposed command, extending the previous sequence and thus ensuring that chosen sequences are incrementally extended, until the round is aborted by a new election. As Perfect Links do not guarantee ordering, both \emph{acceptor} logic must now ensure they only store longer sequences than they already have in the same round. This guarantees that the longest chosen sequence is still a prefix of the accepted sequence, thereby satisfying the \emph{Agreement} property (SC2). Additionally, the leader has to find out what the longest prefix that it has seen $\atom{Accepted}$ (formerly $\atom{Ack}$) messages from a majority, i.e. the longest chosen prefix, before deciding on a new sequence. Only if that prefix is longer than what it has already decided (variable $l_c$), can it issue a $\atom{Decide}$ and only for that prefix.\\
The complete pseudocode can be seen in algorithm \ref{alg:seqpaxos2state}.
\subsubsection{Correctness} By adding Leader Election, the new algorithm fixes the liveness issues of the previous algorithms as discussed at the end of section \ref{sec:paxos}. Now a replica plays the role of a proposer, acceptor, and learner. Since there were no assumptions on role distribution before, making a stronger assumption here does not affect correctness at all. The introduction of the states is simply a consequence of the previous two decisions and introduction of pipelining, and does not cause any issues by itself. The pipelining is an optimisation that is guaranteed to be safe, as \emph{acceptors} that have moved to a new leader already, will ignore $\atom{Accept}$ messages from their old leader that has a lower ballot. As $\atom{Decide}$ messages are only issued for chosen sequences, \textbf{SC2} and \textbf{SC3} are satisfied.

\subsubsection{Performance} The prepare phase takes a single round trip, after which all commands can be ``pipelined''. Pipelining allows many commands to be ``in flight'' in parallel, and for each new command it only takes a single round-trip to get decided locally, as long as the leader remains in place. However, we still send full sequences causing degrading performance with the growth of the log. And, as we have merged the roles now, we are actually also keeping (mostly) redundant sequences in memory, specifically $v_L, v_a$, and $v_d$.

\begin{algorithm}[htp]
\caption{Sequence Paxos 2 -- State}
\label{alg:seqpaxos2state}
\textbf{Implements:} Sequence Consensus\\
\textbf{Requires:} Perfect Link, BLE \\
\textbf{Algorithm:} \\
\HiLi \nl$\Pi$ \tcc*{set of processes}
\HiLi \nl$\id{state} \gets (\atom{follower}, \bot)$\tcc*{role and phase state}
\tcc{Proposer State}
\HiLi\nl$(n_L, v_L) \gets (\bot, \emptyseq)$ \tcc*{Leader's round number and sequence}
\nl$\id{promises}$ $\gets \emptyset$\;
\HiLi \nl$\id{las} \gets [0]^{|\Pi|}$\tcc*{Length of longest accepted sequence per acceptor}
\HiLi \nl$\id{propCmds} \gets \emptyset$\tcc*{A set of commands that need to be appended to the log.}
\HiLi \nl$l_c \gets 0$ \tcc*{Length of longest chosen sequence}
\tcc{Acceptor State}
\nl$n_{prom}\gets 0$ \tcc*{promise not to accept in lower rounds}
\nl$(n_a, v_a) \gets (\bot, \emptyseq)$ \tcc*{round number and sequence accepted}
\tcc{Learner State}
\nl$v_d$ $\gets \emptyseq$ \tcc*{decided sequence}
\end{algorithm}
\addtocounter{algocf}{-1}
\begin{algorithm}[htp]
\caption{Sequence Paxos 2 -- Acceptor\&Learner}
\label{alg:seqpaxos2acc}
\tcc{Acceptor Code}
\nl\Hdl{$\msg{Prepare}{n} \from{p}$}
{
\nl \If{$n_{prom} < n$}{
\nl$n_{prom} \gets n$\;
\nl$\send{\msg{Promise}{n, n_a, v_a}}{p}$\;
}
}
\nl \Hdl{$\msg{Accept}{n, v} \from{p}$}
{
\nl \If{$n_{prom} \leq n$}{
\nl$n_{prom} \gets n$\;
\HiLi\nl$(n_a, v_a) \gets \atom{max}((n, v), (n_a, v_a))$\;
\HiLi\nl$\send{\msg{Accepted}{n, |v_a|}}{p}$\;
}
}
\hrulefill \\
\tcc{Learner Code}
\nl\Hdl{$\msg{Decide}{v}$}
{
\nl \If{$|v_d| < |v|$}{
\nl$v_d \gets v$\;
\nl$\trigger{\msg{Decide}{v_d}}$\;
}
}
\hrulefill \\
\nl\Fn{$\atom{max}((n, v), (n', v'))$}{
\eIf{$n \neq n'$}{
\eIf{$n > n'$}{(n, v)}{(n', v')}
}{
\eIf{$|v| > |v'|$}{(n, v)}{(n', v')}
}
}
\end{algorithm}
\addtocounter{algocf}{-1}
\begin{algorithm}[htp]
\caption{Sequence Paxos 2 -- Proposer}
\label{alg:seqpaxos2prop}
\tcc{Proposer Code}
\nl\Hdl{$\msg{Leader}{L, n}$}
{
\HiLi\eIf{$\id{self} = L \wedge n > n_L \wedge n > n_{prom}$}{
\HiLi\nl$n_L \gets n$\;
\nl$\id{promises}$ $\gets \emptyset$\;
\HiLi \nl$\id{las} \gets [0]^{|\Pi|}$\;
\HiLi \nl$l_c \gets 0$\;
\HiLi\nl$\id{state} \gets (\atom{leader}, \atom{prepare})$\;
\ForEach {$p \in \Pi$}{
\nl$\send{\msg{Prepare}{n_L}}{p}$\;
}
}{
\HiLi\nl$\id{state} \gets (\atom{follower}, \id{state}.2)$\;
}
}
\HiLi\nl\Hdl{$\msg{Propose}{C} \st{\id{state} = (\atom{leader}, \atom{prepare})}$}
{
\HiLi\nl$\id{propCmds} \gets \id{propCmds \cup \{C\}}$\;
}
\HiLi\nl\Hdl{$\msg{Propose}{C} \st{\id{state} = (\atom{leader}, \atom{accept})}$}
{
\HiLi\nl$v_L \gets v_L \oplus C$\;
\HiLi\ForEach {$p \in \Pi$}{
\HiLi\nl$\send{\msg{Accept}{n_L, v_L}}{p}$\;
}
}

\HiLi\nl\Hdl{$\msg{Promise}{n, n', v'} \from{a} \st{n=n_L \wedge \id{state} = (\atom{leader}, \atom{prepare})}$}
{
\nl$\id{promises} \gets \id{promises} \cup \{(a, n', v')\}$\tcc*{add $a$ for acceptor disambiguation}
\nl \If{$|\id{promises}| = \majority{\Pi}$}{
\nl$v \gets \atom{maxValue}(\id{promises})$\tcc*{sequence with the largest $n$, longest if equal}
\HiLi\nl$v_L \gets v \oplus C \inlineforall{C \in \id{propCmds}}$\tcc*{adopt $v$ and append}
\HiLi\nl$\id{propCmds} \gets \emptyset$\;
\HiLi\ForEach {$p \in \Pi$}{
\HiLi\nl$\send{\msg{Accept}{n_L, v_L}}{p}$\;
}
\HiLi\nl$\id{state} \gets (\atom{leader}, \atom{accept})$\;
}
}
\HiLi\nl\Hdl{$\msg{Accepted}{n, l_a} \from{a} \st{n=n_L \wedge \id{state} = (\atom{leader}, \atom{accept})}$}
{
\HiLi\nl$\id{las}[a] \gets \max(l_a, \id{las}[a])$\;
\HiLi\nl$m \gets \max\left\{l \in \N \mid \exists_{M \subseteq \Pi} |M| \geq \majority{\Pi} \wedge \forall_{p\in M} \ l \leq \id{las}[p] \right\}$\tcc*{longest chosen}
\HiLi\nl \If{$m > l_c$}{
\HiLi\nl $l_c \gets m$\;
\HiLi\ForEach {$p \in \Pi$}{
\HiLi\nl$\send{\msg{Decide}{\atom{prefix}(v_L, m)}}{p}$\tcc*{Send the chosen prefix to $p$}
}
}
}
\HiLi\tcc{Drop $\atom{Nack}$ as getting stuck is avoided by BLE re-election}
\end{algorithm}

\clearpage

\subsection{Removing Redundant State}
As all the roles are run by each process now, each role's code actually has access to the state variables of the others. Particularly, $v_L, v_a, v_d$ have significant overlap, since $v_d \prec v_a$ and $v_d \prec v_L$. Thus, \emph{if} we can guarantee that $\atom{Accept}$ messages always arrive before the corresponding $\atom{Decide}$ messages, we can simply replace $v_d$ with a pointer $l_d$ into $v_a$ that marks the decided prefix, i.e. $v_d = \atom{prefix}(v_a, l_d)$. Additionally, we can get rid of $v_L$, if we skip sending $\atom{Prepare}$ messages to and receiving promises from the leader itself, by simply writing the equivalent changes directly into its local state.\\
In order to enable the first optimisation, we are going to strengthen our assumption on the channel abstraction. We will now require \emph{FIFO Perfect Links} that provide message ordering guarantees. This requirement costs us nothing in performance, as before out-of-order commands anyway had to be buffered before being decided. Additionally, it is a simple extension to Perfect Links, achieved by adding sequence numbers and buffering delivery. If our implementation of Perfect Links before was, in fact, TCP\footnote{Note that TCP only fulfils this abstraction during a single session.}, then we get FIFO Perfect Links for free.\\
The complete pseudocode can be seen in algorithm \ref{alg:seqpaxos3state}.

\subsubsection{Correctness} FIFO Perfect Links guarantee that $\atom{Accept}$ messages always arrive before the corresponding $\atom{Decide}$, thus it always holds that for any replica $q$ $v_{d \mbox{ at } q} \prec v_{a \mbox{ at } q}$. To remove $v_L$ notice that the leader has access to its own $v_a$. When a process becomes leader it can update its local $v_a$ directly. This is guaranteed to be accepted, since $n_L$ is guaranteed to be higher or equal to the current $n_{prom}$.

\subsubsection{Performance} Apart from saving one message per command (the self-message), these optimisations have mostly reduced the memory footprint of the algorithm. Instead of storing three sequences, we now store only one sequence and one new pointer/index.

\begin{algorithm}[htp]
\caption{Sequence Paxos 3 -- State}
\label{alg:seqpaxos3state}
\textbf{Implements:} Sequence Consensus\\
\textbf{Requires:} FIFO Perfect Link, BLE \\
\textbf{Algorithm:} \\
\nl$\Pi$ \tcc*{set of processes}
\HiLi\nl$\Pi_o \gets \Pi - \{\id{self}\}$\;
\nl$\id{state} \gets (\atom{follower}, \bot)$\tcc*{role and phase state}
\tcc{Proposer State}
\HiLi\nl$n_L \gets 0$ \tcc*{leader's round number}
\nl$\id{promises}$ $\gets \emptyset$\;
\nl$\id{las} \gets [0]^{|\Pi|}$\tcc*{length of longest accepted sequence per acceptor}
\nl$\id{propCmds} \gets \emptyset$\tcc*{set of commands that need to be appended to the log}
\nl$l_c \gets 0$ \tcc*{length of longest chosen sequence}
\tcc{Acceptor State}
\nl$n_{prom}\gets 0$ \tcc*{promise not to accept in lower rounds}
\nl$(n_a, v_a) \gets (\bot, \emptyseq)$ \tcc*{round number and sequence accepted}
\tcc{Learner State}
\HiLi\nl$l_d$ $\gets 0$ \tcc*{length of the decided sequence}
\end{algorithm}
\addtocounter{algocf}{-1}
\begin{algorithm}[htp]
\caption{Sequence Paxos 3 -- Acceptor\&Learner}
\label{alg:seqpaxos3acc}
\tcc{Acceptor Code}
\nl\Hdl{$\msg{Prepare}{n} \from{p}$}
{
\nl \If{$n_{prom} < n$}{
\nl$n_{prom} \gets n$\;
\nl$\send{\msg{Promise}{n, n_a, v_a}}{p}$\;
}
}
\nl \Hdl{$\msg{Accept}{n, v} \from{p}$}
{
\nl \If{$n_{prom} \leq n$}{
\nl$n_{prom} \gets n$\;
\nl$(n_a, v_a) \gets \atom{max}((n, v), (n_a, v_a))$\;
\nl$\send{\msg{Accepted}{n, |v_a|}}{p}$\;
}
}
\hrulefill \\
\tcc{Learner Code}
\HiLi\nl\Hdl{$\msg{Decide}{v, n} \st{n=n_{prom}}$}
{
\HiLi\nl \If{$l_d < |v|$}{
\HiLi\nl$l_d \gets |v|$\;
\HiLi\nl$\trigger{\msg{Decide}{\atom{prefix}(v_a, l_d)}}$\;
}
}
\hrulefill \\
\nl\Fn{$\atom{max}((n, v), (n', v'))$}{
\eIf{$n \neq n'$}{
\eIf{$n > n'$}{(n, v)}{(n', v')}
}{
\eIf{$|v| > |v'|$}{(n, v)}{(n', v')}
}
}
\end{algorithm}
\addtocounter{algocf}{-1}
\begin{algorithm}[htp]
\caption{Sequence Paxos 3 -- Proposer}
\label{alg:seqpaxos3prop}
\tcc{Proposer Code}
\nl\Hdl{$\msg{Leader}{L, n}$}
{
\eIf{$\id{self} = L \wedge n > n_L \wedge n > n_{prom}$}{
\HiLi\nl$(n_L, n_{prom}) \gets (n, n)$\;
\HiLi\nl$\id{promises}$ $\gets \{ (\id{self}, n_a, v_a) \}$\;
\nl$\id{las} \gets [0]^{|\Pi|}$\;
\nl$l_c \gets 0$\;
\nl$\id{state} \gets (\atom{leader}, \atom{prepare})$\;
\HiLi\ForEach {$p \in \Pi_o$}{
\nl$\send{\msg{Prepare}{n_L}}{p}$\;
}
}{
\nl$\id{state} \gets (\atom{follower}, \id{state}.2)$\;
}
}
\nl\Hdl{$\msg{Propose}{C} \st{\id{state} = (\atom{leader}, \atom{prepare})}$}
{
\nl$\id{propCmds} \gets \id{propCmds \cup \{C\}}$\;
}
\nl\Hdl{$\msg{Propose}{C} \st{\id{state} = (\atom{leader}, \atom{accept})}$}
{
\HiLi\nl$v_a \gets v_a \oplus C$\;
\HiLi\nl$las[\id{self}] \gets |v_a|$\;
\HiLi\ForEach {$p \in \Pi_o$}{
\HiLi\nl$\send{\msg{Accept}{n_L, v_a}}{p}$\;
}
}

\nl\Hdl{$\msg{Promise}{n, n', v'} \from{a} \st{n=n_L \wedge \id{state} = (\atom{leader}, \atom{prepare})}$}
{
\nl$\id{promises} \gets \id{promises} \cup \{(a, n', v')\}$\tcc*{add $a$ for acceptor disambiguation}
\nl \If{$|\id{promises}| = \majority{\Pi}$}{
\nl$v \gets \atom{maxValue}(\id{promises})$\tcc*{sequence with the largest $n$, longest if equal}
\HiLi\nl$v_a \gets v \oplus C \inlineforall{C \in \id{propCmds}}$\tcc*{adopt $v$ and append}
\nl$\id{propCmds} \gets \emptyset$\;
\HiLi\nl$las[\id{self}] \gets |v_a|$\;
\HiLi\ForEach {$p \in \Pi_o$}{
\HiLi\nl$\send{\msg{Accept}{n_L, v_a}}{p}$\;
}
\nl$\id{state} \gets (\atom{leader}, \atom{accept})$\;
}
}
\nl\Hdl{$\msg{Accepted}{n, l_a} \from{a} \st{n=n_L \wedge \id{state} = (\atom{leader}, \atom{accept})}$}
{
\HiLi\nl$\id{las}[a] \gets l_a$\;
\HiLi\nl$M \gets \{p \in \Pi \mid \id{las}[p] \neq \bot \wedge \id{las}[p] \geq l_a\}$\tcc*{support set for $l_a$}
\HiLi\nl \If{$l_a > l_c \wedge |M| \geq \majority{\Pi}$}{
\HiLi\nl $l_c \gets l_a$\;
\ForEach {$p \in \Pi$}{
\HiLi\nl$\send{\msg{Decide}{\atom{prefix}(v_a, l_a)}, n_L}{p}$\tcc*{send chosen prefix}
}
}
}
\end{algorithm}

\clearpage

\subsection{Avoid Sending Sequences}
The current algorithm still sends full sequences with every single message. As discussed before, these sequences will overlap in most positions and at most differ slightly at the end. Furthermore, in the $\msg{Decide}{v, n}$ message we are not even using $v$ at all, but only its length, as we already have access to $v_a$ and we know $v \prec v_a$. Thus we will simply replace $v$ with $|v| = l_c$ in the $\atom{Decide}$ message.

In addition to this simple change, we also want to trim the data we send in the prepare phase as much as possible. To preserve correctness a leader \emph{must} extend a sequence that contains $v_d$ at the end of the prepare phase. But both the leader and some acceptors may be ahead or behind $v_d$ with their $v_a$ during the prepare phase. We want to synchronise everyone by sending as little data as possible. As a majority of acceptors had at least the chosen sequence before the leader change, any majority now must include at least a single acceptor that still knows it (as a majority may not fail). If the leader tells everyone what its decided sequence is (i.e. add $l_d$ to $\atom{Prepare}$ messages), the acceptors can either catch up the leader (if \emph{it} is behind) or ask to be caught up themselves, if \emph{they} are behind. That is, for a leader $L$ at every acceptor $a$, if $l_{d \mbox{ at } a} > l_{d \mbox{ at } L}$ then $a$ will send $\atom{suffix}(v_a, l_{d \mbox{ at } L})$ as part of its $\atom{Promise}$ to catch up the leader. Otherwise it will send $\emptyseq$ and wait for the leader to catch it up with the first $\atom{Accept}$ message. After collecting a majority, the leader will adopt the \emph{max suffix}, that is the suffix with highest round number or, if the round numbers are equal, the longest. It will then append it to its decided sequence ($v_d= \atom{prefix}(v_a, l_d)$) and append all the commands it wanted to propose. That value will becomes its new $v_a$ and that is also the sequence it will impose on all followers.

At this point only the $\atom{Accept}$ messages continue to send full sequences. In order to avoid this, we split off the first $\atom{Accept}$ at the end of the \emph{prepare phase} into a new message $\atom{AcceptSync}$, as this message has a different purpose than the $\atom{Accept}$ messages being generated from proposals. The purpose of the $\atom{AcceptSync}$ message is get every replica to the same state of the leader, by sending them only as much data as they really need. The purpose of the $\atom{Accept}$ message is to append a single new command to $v_a$. If the replicas are in sync with the leader (i.e. have been part of the prepare phase), then simply sending this command alone will suffice, considering that the FIFO Perfect Links will preserve the sequence order. In order to avoid sending the whole sequence in $\atom{AcceptSync}$, replicas have to inform the leader of their $l_d$ in the $\atom{Prepare}$ message, which the leader stores in a new map $\id{lds}$ from each acceptor to the length of their decided sequence. Instead of sending $v_a$ the leader then sends $\atom{suffix}(v_a, lds[a])$ to every acceptor $a$. 

As only a majority of processes have responded when the leader transitions into the \emph{accept phase}, we must add extra message handlers to catch up late replicas. Particularly, we must \emph{not} send $\atom{Accept}$ messages to a replica until we have processed its $\atom{Promise}$, as those replicas are out of sync. The same reasoning goes for $\atom{Decide}$ messages for late replicas. That is, if a leader has already issued decides during its \emph{accept phase} when it gets a promise from a late replica, it must send both an up-to-date $\atom{AcceptSync}$, immediately followed by a $\atom{Decide}$ for the currently longest chosen sequence $l_c$.\\
The complete pseudocode for these optimisations can be seen in algorithm \ref{alg:seqpaxos4state}.

\subsubsection{Correctness} Sending $|v|=l_c$ in $\atom{Decide}$ messages instead of $v$, is a trivial change, as only $|v|$ was used before anyway. Sending only diffs instead of full sequences also does not affect the correctness as long as the correct diffs are being sent. The renaming of the first $\atom{Accept}$ message to $\atom{AcceptSync}$ itself has no effect, it simply reflects the usage more accurately. It would also be possible to infer this information (whether a message is $\atom{Accept}$ or $\atom{AcceptSync}$) from context, but that is less readable. The crucial part here is to use FIFO Perfect Links and $\id{lds}$ to make sure that an $\atom{AcceptSync}$ always arrives before the first $\atom{Accept}$ in that round.

\subsubsection{Performance} Not sending full sequences anymore is a huge improvement in performance, especially as it finally removes the performance degradation issue with large sequences. At this point, we made leader changes as cheap as we can, and during stable periods we have full pipelining for single commands. If this leads to bad performance due to network packets being too small, it could easily be extended again to do some kind of batching. Although whether this should happen within \emph{Sequence Paxos} or in an external management component is an implementation decision.

\begin{algorithm}[htp]
\caption{Sequence Paxos 4 -- State}
\label{alg:seqpaxos4state}
\textbf{Implements:} Sequence Consensus\\
\textbf{Requires:} FIFO Perfect Link, BLE \\
\textbf{Algorithm:} \\
\nl$\Pi$ \tcc*{set of processes}
\nl$\Pi_o \gets \Pi - \{\id{self}\}$\;
\nl$\id{state} \gets (\atom{follower}, \bot)$\tcc*{role and phase state}
\tcc{Proposer State}
\nl$n_L \gets 0$ \tcc*{leader's round number}
\nl$\id{promises}$ $\gets \emptyset$\;
\nl$\id{las} \gets [0]^{|\Pi|}$\tcc*{length of longest accepted sequence per acceptor}
\HiLi\nl$\id{lds} \gets [\bot]^{|\Pi|}$\tcc*{length of longest known decided sequence per acceptor}
\nl$\id{propCmds} \gets \emptyset$\tcc*{set of commands that need to be appended to the log}
\nl$l_c \gets 0$ \tcc*{length of longest chosen sequence}
\tcc{Acceptor State}
\nl$n_{prom}\gets 0$ \tcc*{promise not to accept in lower rounds}
\HiLi\nl$(n_a, v_a) \gets (0, \emptyseq)$ \tcc*{round number and sequence accepted}
\tcc{Learner State}
\nl$l_d$ $\gets 0$ \tcc*{length of the decided sequence}
\end{algorithm}
\addtocounter{algocf}{-1}
\begin{algorithm}[htp]
\caption{Sequence Paxos 4 -- Acceptor\&Learner}
\label{alg:seqpaxos4acc}
\tcc{Acceptor Code}
\HiLi\nl\Hdl{$\msg{Prepare}{n, \id{ld}} \from{p}$}
{
\nl \If{$n_{prom} < n$}{
\nl$n_{prom} \gets n$\;
\HiLi\nl$\id{state} \gets (\atom{follower}, \atom{prepare})$\;
\HiLi\nl$\id{suffix} \gets \atom{suffix}(v_a, \id{ld})$\;
\HiLi\nl$\send{\msg{Promise}{n, n_a, \id{suffix}, l_d}}{p}$\;
}
}
\HiLi\nl \Hdl{$\msg{AcceptSync}{n, \id{suffix}, \id{ld}} \from{p} \st{\id{state} = (\atom{follower}, \atom{prepare})}$}
{
\HiLi\nl \If{$n_{prom} = n$}{
\HiLi\nl$n_a \gets n$\;
\HiLi\nl$v_a \gets \atom{prefix}(v_a, \id{ld}) \append \id{suffix}$\;
\HiLi\nl$\id{state} \gets (\atom{follower}, \atom{accept})$\;
\nl$\send{\msg{Accepted}{n, |v_a|}}{p}$\;
}
}
\HiLi\nl \Hdl{$\msg{Accept}{n, C} \from{p} \st{\id{state} = (\atom{follower}, \atom{accept})}$}
{
\HiLi\nl \If{$n_{prom} = n$}{
\HiLi\nl$v_a \gets v_a \oplus C$\;
\nl$\send{\msg{Accepted}{n, |v_a|}}{p}$\;
}
}
\hrulefill \\
\tcc{Learner Code}
\HiLi\nl\Hdl{$\msg{Decide}{l, n} \st{n=n_{prom}}$}
{
\HiLi\nl \If{$l_d < l$}{
\HiLi\nl$l_d \gets l$\;
\nl$\trigger{\msg{Decide}{\atom{prefix}(v_a, l_d)}}$\;
}
}
\hrulefill \\
\nl\Fn{$\atom{max}((n, v), (n', v'))$}{
\eIf{$n \neq n'$}{
\eIf{$n > n'$}{(n, v)}{(n', v')}
}{
\eIf{$|v| > |v'|$}{(n, v)}{(n', v')}
}
}
\end{algorithm}
\addtocounter{algocf}{-1}
\begin{algorithm}[htp]
\caption{Sequence Paxos 4 -- Proposer}
\label{alg:seqpaxos4prop}
\tcc{Proposer Code}
\nl\Hdl{$\msg{Leader}{L, n}$}
{
\eIf{$\id{self} = L \wedge n > n_L \wedge n > n_{prom}$}{
\nl$(n_L, n_{prom}) \gets (n, n)$\;
\HiLi\nl$\id{promises}$ $\gets \{ (\id{self}, n_a, \atom{suffix}(v_a, l_d)) \}$\;
\nl$\id{las} \gets [0]^{|\Pi|}$\;
\HiLi\nl$\id{lds} \gets [\bot]^{|\Pi|}; \id{lds}[\id{self}] \gets l_d$\;
\nl$l_c \gets 0$\;
\nl$\id{state} \gets (\atom{leader}, \atom{prepare})$\;
\ForEach {$p \in \Pi_o$}{
\HiLi\nl$\send{\msg{Prepare}{n_L, l_d}}{p}$\;
}
}{
\nl$\id{state} \gets (\atom{follower}, \id{state}.2)$\;
}
}
\nl\Hdl{$\msg{Propose}{C} \st{\id{state} = (\atom{leader}, \atom{prepare})}$}
{
\nl$\id{propCmds} \gets \id{propCmds \cup \{C\}}$\;
}
\nl\Hdl{$\msg{Propose}{C} \st{\id{state} = (\atom{leader}, \atom{accept})}$}
{
\nl$v_a \gets v_a \oplus C$\;
\nl$las[\id{self}] \gets |v_a|$\;
\HiLi\ForEach {$p \in \{p\in\Pi_0 \mid \id{lds}[p] \neq \bot\}$}{
\HiLi\nl$\send{\msg{Accept}{n_L, C}}{p}$\;
}
}

\HiLi\nl\Hdl{$\msg{Promise}{n, n', \id{suffix}_a, ld_a} \from{a} \st{n=n_L \wedge \id{state} = (\atom{leader}, \atom{prepare})}$}
{
\HiLi\nl$\id{promises} \gets \id{promises} \cup \{(a, n', \id{suffix}_a)\}$\;
\HiLi\nl$\id{lds}[a] \gets ld_a$\;
\nl \If{$|\id{promises}| = \majority{\Pi}$}{
\nl$\id{suffix} \gets \atom{maxValue}(\id{promises})$\tcc*{suffix with max $n$, longest if equal}
\tcc{adopt $v_d \append \id{suffix}$ and append commands}
\HiLi\nl$v_a \gets \atom{prefix}(v_a, l_d) \append \id{suffix} \oplus C \inlineforall{C \in \id{propCmds}}$\;
\nl$\id{propCmds} \gets \emptyset$\;
\nl$las[\id{self}] \gets |v_a|$\;
\nl$\id{state} \gets (\atom{leader}, \atom{accept})$\;
\HiLi\ForEach {$p \in \{p\in\Pi_0 \mid \id{lds}[p] \neq \bot\}$}{
\HiLi\nl$\send{\msg{AcceptSync}{n_L, \atom{suffix}(v_a, \id{lds}[p]), \id{lds}[p]}}{p}$\;
}
}
}
\HiLi\nl\Hdl{$\msg{Promise}{n, n', \id{suffix}_a, ld_a} \from{a} \st{n=n_L \wedge \id{state} = (\atom{leader}, \atom{accept})}$}
{
\HiLi\nl$\id{lds}[a] \gets ld_a$\;
\HiLi\nl$\send{\msg{AcceptSync}{n_L, \atom{suffix}(v_a, \id{lds}[a]), \id{lds}[a]}}{p}$\;
\HiLi\nl\If{$l_c \neq 0$}{
\HiLi\nl$\send{\msg{Decide}{l_c, n_L}}{a}$\tcc*{also inform what got decided already}
}
}

\nl\Hdl{$\msg{Accepted}{n, l_a} \from{a} \st{n=n_L \wedge \id{state} = (\atom{leader}, \atom{accept})}$}
{
\nl$\id{las}[a] \gets l_a$\;
\nl$M \gets \{p \in \Pi \mid \id{las}[p] \neq \bot \wedge \id{las}[p] \geq l_a\}$\tcc*{support set for $l_a$}
\nl \If{$l_a > l_c \wedge |M| \geq \majority{\Pi}$}{
\nl $l_c \gets l_a$\;
\HiLi\ForEach {$p \in \{p\in\Pi_0 \mid \id{lds}[p] \neq \bot\}$}{
\HiLi\nl$\send{\msg{Decide}{l_c, n_L}}{p}$\tcc*{send length of chosen sequence}
}
}
}
\end{algorithm}
\clearpage

\subsection{Final Optimisations}
At this point we have a pretty efficient algorithm and all that is left is some minor optimisations and convenience fixes. One thing to notice is that the $\atom{Promise}$ at acceptor $a$ currently sends a suffix to leader $L$ even when $n_{a \mbox{ at } L} > n_{a \mbox{ at } a}$, although $\atom{maxValue}(\id{promises})$ at the leader would never pick that suffix to be adopted. Na\"{i}vely, we might think this situation would never occur, as a process that lags behind one round, probably should not have a longer $v_a$ anyway. But this is not true. Consider a leader $L_1$ in round 1 getting disconnected from the rest of the group, but still extending its sequence locally. In the mean time a new leader $L_2$ is elected in round 2 with the remaining nodes, that $L_1$ never knows about. Shortly after, the connection stabilises and a new leader $L_3$ is elected in round 3 with $L_1$ part of the majority. If $L_2$ didn't have much time to add commands, it might be that case that $|v_{a \mbox{ at } L_1}| > |v_{a \mbox{ at } L_3}|$, even though $n_{a \mbox{ at } L_3} > n_{a \mbox{ at } L_1}$. While this is an edge case, $L_1$ could have millions of messages in its local log, that will never make it into the new $v_a$ as they conflict with the decisions made in its absence. Thus sending them is unnecessary. To avoid this situation the leader will add its local $n_a$ to every $\atom{Prepare}$ message, which an acceptor will check before calculating the suffix to send in the $\atom{Promise}$.\\
The other optimisation in this section is meant to reduce redundancy between the replicated log service of \emph{Sequence Consensus} and the state machine on top of it executing commands. Currently a $\atom{Decide}$ event includes the full log. However, the vast majority if not all state machine implementation will not start with the \emph{initial state} and apply the full log every time a decision is made. Instead they will only apply the new commands one at a time to the previously stored state. As we are already keeping track of previously decided prefixes in \emph{Sequence Paxos}, it is redundant to keep track of this again in the state machine. To avoid this we will make another small alteration to the \emph{Sequence Paxos} interface, in which we will only decide a single command at a time, with the semantic meaning of this being appended to the previously decided commands (or simply immediately consumed by the state machine).\\
The complete pseudocode for the final version can be seen in algorithm \ref{alg:seqpaxosfinalstate}.

\subsubsection{Correctness} Both changes are minor and trivially correct, with the caveat that the latter change technically implements a different interface.

\subsubsection{Performance} The performance is the same as in the previous version, with the exception of the edge case described above where $n_{a \mbox{ at } L} > n_{a \mbox{ at } a}$ for some leader $L$ and acceptor $a$.

\begin{algorithm}[hbtp]
\caption{Sequence Paxos Final -- State}
\label{alg:seqpaxosfinalstate}
\textbf{Implements:} Sequence Consensus\\
\textbf{Requires:} FIFO Perfect Link, BLE \\
\textbf{Algorithm:} \\
\nl$\Pi$ \tcc*{set of processes}
\nl$\Pi_o \gets \Pi - \{\id{self}\}$\;
\nl$\id{state} \gets (\atom{follower}, \bot)$\tcc*{role and phase state}
\tcc{Proposer State}
\nl$n_L \gets 0$ \tcc*{leader's round number}
\nl$\id{promises}$ $\gets \emptyset$\;
\nl$\id{las} \gets [0]^{|\Pi|}$\tcc*{length of longest accepted sequence per acceptor}
\nl$\id{lds} \gets [\bot]^{|\Pi|}$\tcc*{length of longest known decided sequence per acceptor}
\nl$\id{propCmds} \gets \emptyset$\tcc*{set of commands that need to be appended to the log}
\nl$l_c \gets 0$ \tcc*{length of longest chosen sequence}
\tcc{Acceptor State}
\nl$n_{prom}\gets 0$ \tcc*{promise not to accept in lower rounds}
\nl$(n_a, v_a) \gets (0, \emptyseq)$ \tcc*{round number and sequence accepted}
\tcc{Learner State}
\nl$l_d$ $\gets 0$ \tcc*{length of the decided sequence}
\end{algorithm}
\addtocounter{algocf}{-1}
\begin{algorithm}[htp]
\caption{Sequence Paxos Final -- Acceptor\&Learner}
\label{alg:seqpaxosfinalacc}
\tcc{Acceptor Code}
\HiLi\nl\Hdl{$\msg{Prepare}{n, \id{ld}, \id{na}_L} \from{p}$}
{
\nl \If{$n_{prom} < n$}{
\nl$n_{prom} \gets n$\;
\nl$\id{state} \gets (\atom{follower}, \atom{prepare})$\;
\HiLi\nl$\id{suffix} \gets \inlineif{n_a \geq \id{na}_L}{\atom{suffix}(v_a, \id{ld})}{\emptyseq}$\;
\nl$\send{\msg{Promise}{n, n_a, \id{suffix}, l_d}}{p}$\;
}
}
\nl \Hdl{$\msg{AcceptSync}{n, \id{suffix}, \id{ld}} \from{p} \st{\id{state} = (\atom{follower}, \atom{prepare})}$}
{
\nl \If{$n_{prom} = n$}{
\nl$n_a \gets n$\;
\nl$v_a \gets \atom{prefix}(v_a, \id{ld}) \append \id{suffix}$\;
\nl$\id{state} \gets (\atom{follower}, \atom{accept})$\;
\nl$\send{\msg{Accepted}{n, |v_a|}}{p}$\;
}
}
\nl \Hdl{$\msg{Accept}{n, C} \from{p} \st{\id{state} = (\atom{follower}, \atom{accept})}$}
{
\nl \If{$n_{prom} = n$}{
\nl$v_a \gets v_a \oplus C$\;
\nl$\send{\msg{Accepted}{n, |v_a|}}{p}$\;
}
}
\hrulefill \\
\tcc{Learner Code}
\nl\Hdl{$\msg{Decide}{l, n} \st{n=n_{prom}}$}
{
\HiLi\nl \While{$l_d < l$}{
\HiLi\nl$\trigger{\msg{Decide}{v_a[l_d]}}$\tcc*{assuming 0-based indexing}
\HiLi\nl$l_d \gets l_d + 1$\;
}
}
\end{algorithm}
\addtocounter{algocf}{-1}
\begin{algorithm}[htp]
\caption{Sequence Paxos Final -- Proposer}
\label{alg:seqpaxosfinalprop}
\tcc{Proposer Code}
\nl\Hdl{$\msg{Leader}{L, n}$}
{
\eIf{$\id{self} = L \wedge n > n_L \wedge n > n_{prom}$}{
\nl$(n_L, n_{prom}) \gets (n, n)$\;
\nl$\id{promises}$ $\gets \{ (\id{self}, n_a, \atom{suffix}(v_a, l_d)) \}$\;
\nl$\id{las} \gets [0]^{|\Pi|}$\;
\nl$\id{lds} \gets [\bot]^{|\Pi|}; \id{lds}[\id{self}] \gets l_d$\;
\nl$l_c \gets 0$\;
\nl$\id{state} \gets (\atom{leader}, \atom{prepare})$\;
\ForEach {$p \in \Pi_o$}{
\HiLi\nl$\send{\msg{Prepare}{n_L, l_d, n_a}}{p}$\;
}
}{
\nl$\id{state} \gets (\atom{follower}, \id{state}.2)$\;
}
}
\nl\Hdl{$\msg{Propose}{C} \st{\id{state} = (\atom{leader}, \atom{prepare})}$}
{
\nl$\id{propCmds} \gets \id{propCmds \cup \{C\}}$\;
}
\nl\Hdl{$\msg{Propose}{C} \st{\id{state} = (\atom{leader}, \atom{accept})}$}
{
\nl$v_a \gets v_a \oplus C$\;
\nl$las[\id{self}] \gets |v_a|$\;
\ForEach {$p \in \{p\in\Pi_0 \mid \id{lds}[p] \neq \bot\}$}{
\nl$\send{\msg{Accept}{n_L, C}}{p}$\;
}
}

\nl\Hdl{$\msg{Promise}{n, n', \id{suffix}_a, ld_a} \from{a} \st{n=n_L \wedge \id{state} = (\atom{leader}, \atom{prepare})}$}
{
\nl$\id{promises} \gets \id{promises} \cup \{(a, n', \id{suffix}_a)\}$\;
\nl$\id{lds}[a] \gets ld_a$\;
\nl \If{$|\id{promises}| = \majority{\Pi}$}{
\nl$\id{suffix} \gets \atom{maxValue}(\id{promises})$\tcc*{suffix with max $n$, longest if equal}
\tcc{adopt $v_d \append \id{suffix}$ and append commands}
\nl$v_a \gets \atom{prefix}(v_a, l_d) \append \id{suffix} \oplus C \inlineforall{C \in \id{propCmds}}$\;
\nl$\id{propCmds} \gets \emptyset$\;
\nl$las[\id{self}] \gets |v_a|$\;
\nl$\id{state} \gets (\atom{leader}, \atom{accept})$\;
\ForEach {$p \in \{p\in\Pi_0 \mid \id{lds}[p] \neq \bot\}$}{
\nl$\send{\msg{AcceptSync}{n_L, \atom{suffix}(v_a, \id{lds}[p]), \id{lds}[p]}}{p}$\;
}
}
}
\nl\Hdl{$\msg{Promise}{n, n', \id{suffix}_a, ld_a} \from{a} \st{n=n_L \wedge \id{state} = (\atom{leader}, \atom{accept})}$}
{
\nl$\id{lds}[a] \gets ld_a$\;
\nl$\send{\msg{AcceptSync}{n_L, \atom{suffix}(v_a, \id{lds}[a]), \id{lds}[a]}}{p}$\;
\nl\If{$l_c \neq 0$}{
\nl$\send{\msg{Decide}{l_c, n_L}}{a}$\tcc*{also inform what got decided already}
}
}

\nl\Hdl{$\msg{Accepted}{n, l_a} \from{a} \st{n=n_L \wedge \id{state} = (\atom{leader}, \atom{accept})}$}
{
\nl$\id{las}[a] \gets l_a$\;
\nl$M \gets \{p \in \Pi \mid \id{las}[p] \neq \bot \wedge \id{las}[p] \geq l_a\}$\tcc*{support set for $l_a$}
\nl \If{$l_a > l_c \wedge |M| \geq \majority{\Pi}$}{
\nl $l_c \gets l_a$\;
\ForEach {$p \in \{p\in\Pi_0 \mid \id{lds}[p] \neq \bot\}$}{
\nl$\send{\msg{Decide}{l_c, n_L}}{p}$\tcc*{send length of chosen sequence}
}
}
}
\end{algorithm}
\clearpage

\section{Fail-Recovery}
\label{sec:recovery}
At his point we have a fairly efficient algorithm as long as a majority never crashes. This limitation arises from our use of the fail-stop model, so far, where correctness means never crashing. For any long-running service an assumption of never losing a cumulative majority becomes impractical, as even small failure probabilities accumulate over time. We will treat this issue in two distinct steps: This section will deal with transient failures, where the OS process crashes and then is started again or a physical host reboots, for example. In section \ref{sec:reconf} we will describe how to deal with permanent process failures by reconfiguring the members of the replication group.\\
In the \emph{fail-recovery} model, a process is considered \emph{correct} if it crashes and consequently recovers a finite number of times in an execution. During the crash a process may lose all its state (called \emph{amnesia}) and an arbitrary suffix of the most recent messages (\emph{omission} failures). However, a node may store some of its state in a \emph{persistent} manner, which can be loaded during recovery. In reality this translates to storage on disk, for example. As persistent storage typically has a performance impact, we want to store only the minimum necessary state in this manner.

\subsection{Sequence Paxos Recovery}
\label{ssec:failrec}
In order to augment algorithm \ref{alg:seqpaxosfinalstate} for the fail-recovery model, we introduce a new $\id{state}$ called $\atom{recover}$, which is automatically entered, when the system detects that it has state from a previous run available. The following variables need to be stored in persistent storage and loaded during recovery: $n_{\id{prom}}, n_a, v_a, l_d$. We will also require our \emph{BLE} implementation to start with $\id{ballot_{\id{max}}} = n_{\id{prom}}$, reusing the already stored variable from \emph{Sequence Paxos} for efficiency.\\
During recovery a process $p$ starts with $\id{state}=(\atom{follower}, \atom{recover})$ and restores all the persistent variables. It then waits for a $\msg{Leader}{L, n}$ message, ignoring all other messages.
\setcounter{case}{0}
\begin{case}[$p=L$] In this case $p$ has been elected leader and should simply run a \emph{prepare phase} as normal. Whatever state and messages it has missed while crashed will be synced up during the \emph{prepare phase}.
\end{case}
\begin{case}[$p\neq L$] In this case $p$ is a follower, and there is some other leader actively sending $\atom{Accept}$ messages already. It is unclear whether $p$ had completed the \emph{prepare phase} before crashing, but in any case it needs to sync up again with the other nodes before it can handle any $\atom{Accept}$ messages. To facilitate this with minimal code changes, we will introduce a new message called $\atom{PrepareReq}$ which $p$ will send to $L$. Then $p$ will wait for $L$ to send it a $\atom{Prepare}$ message. From this point the algorithm proceeds as if $p$ was simply a late process without any further changes. If $L$ was stuck in the \emph{prepare phase} due to a missing majority without $p$, resending $\atom{Prepare}$ to $p$ will help it make progress. If $L$ already moved to the \emph{accept} phase, it will know how to catch up $p$ with an $\atom{AcceptSync}$ message.
\end{case}

The complete pseudocode for this version can be seen in algorithm \ref{alg:seqpaxosrecoverystate}.

\subsubsection{Correctness} All the state variables apart from $n_{\id{prom}}, n_a, v_a, l_d$ is going to be overwritten during the \emph{prepare phase}, which we are running after recovery anyway, so storing it would be redundant. As the purpose of the \emph{prepare phase} is to get all the replicas in sync, it can easily be seen that the proposed behaviour will counter any \emph{amnesia} or \emph{omission} that occurred during the crash at $p$.

\begin{algorithm}[hbtp]
\caption{Sequence Paxos Fail-Recovery -- State}
\label{alg:seqpaxosrecoverystate}
\textbf{Implements:} Sequence Consensus\\
\textbf{Requires:} FIFO Perfect Link, BLE \\
\textbf{Algorithm:} \\
\nl$\Pi$ \tcc*{set of processes}
\nl$\Pi_o \gets \Pi - \{\id{self}\}$\;
\nl$\id{state} \gets (\atom{follower}, \bot)$\tcc*{role and phase state}
\tcc{Proposer State}
\nl$n_L \gets 0$ \tcc*{leader's round number}
\nl$\id{promises}$ $\gets \emptyset$\;
\nl$\id{las} \gets [0]^{|\Pi|}$\tcc*{length of longest accepted sequence per acceptor}
\nl$\id{lds} \gets [\bot]^{|\Pi|}$\tcc*{length of longest known decided sequence per acceptor}
\nl$\id{propCmds} \gets \emptyset$\tcc*{set of commands that need to be appended to the log}
\nl$l_c \gets 0$ \tcc*{length of longest chosen sequence}
\tcc{Acceptor State}
\HiLi\nl$\persist{n_{prom}}\gets 0$ \tcc*{promise not to accept in lower rounds}
\HiLi\nl$\persist{(n_a, v_a)} \gets (0, \emptyseq)$ \tcc*{round number and sequence accepted}
\tcc{Learner State}
\HiLi\nl$\persist{l_d} \gets 0$ \tcc*{length of the decided sequence}
\end{algorithm}
\addtocounter{algocf}{-1}
\begin{algorithm}[htp]
\caption{Sequence Paxos Fail-Recovery -- Part 1}
\label{alg:seqpaxosrecovery1}
\tcc{Proposer Code}
\nl\Hdl{$\msg{Leader}{L, n}$}
{
\eIf{$\id{self} = L \wedge n > n_L \wedge n > n_{prom}$}{
\nl$(n_L, n_{prom}) \gets (n, n)$\;
\nl$\id{promises}$ $\gets \{ (\id{self}, n_a, \atom{suffix}(v_a, l_d)) \}$\;
\nl$\id{las} \gets [0]^{|\Pi|}$\;
\nl$\id{lds} \gets [\bot]^{|\Pi|}; \id{lds}[\id{self}] \gets l_d$\;
\nl$l_c \gets 0$\;
\nl$\id{state} \gets (\atom{leader}, \atom{prepare})$\;
\ForEach {$p \in \Pi_o$}{
\nl$\send{\msg{Prepare}{n_L, l_d, n_a}}{p}$\;
}
}{
\HiLi\eIf{$\id{state}=(\_, \atom{recover})$}{
\HiLi\nl$\send{\msgatom{PrepareReq}}{L}$\;
}{
\nl$\id{state} \gets (\atom{follower}, \id{state}.2)$\;
}
}
}

\HiLi\nl\Hdl{$\msgatom{PrepareReq} \from{a} \st{\id{state} = (\atom{leader}, \_)}$}
{
\HiLi\nl$\send{\msg{Prepare}{n_L, l_d, n_a}}{a}$\;
}

\nl\Hdl{$\msg{Propose}{C} \st{\id{state} = (\atom{leader}, \atom{prepare})}$}
{
\nl$\id{propCmds} \gets \id{propCmds \cup \{C\}}$\;
}
\nl\Hdl{$\msg{Propose}{C} \st{\id{state} = (\atom{leader}, \atom{accept})}$}
{
\nl$v_a \gets v_a \oplus C$\;
\nl$las[\id{self}] \gets |v_a|$\;
\ForEach {$p \in \{p\in\Pi_0 \mid \id{lds}[p] \neq \bot\}$}{
\nl$\send{\msg{Accept}{n_L, C}}{p}$\;
}
}

\nl\Hdl{$\msg{Promise}{n, n', \id{suffix}_a, ld_a} \from{a} \st{n=n_L \wedge \id{state} = (\atom{leader}, \atom{prepare})}$}
{
\nl$\id{promises} \gets \id{promises} \cup \{(a, n', \id{suffix}_a)\}$\;
\nl$\id{lds}[a] \gets ld_a$\;
\nl \If{$|\id{promises}| = \majority{\Pi}$}{
\nl$\id{suffix} \gets \atom{maxValue}(\id{promises})$\tcc*{suffix with max $n$, longest if equal}
\tcc{adopt $v_d \append \id{suffix}$ and append commands}
\nl$v_a \gets \atom{prefix}(v_a, l_d) \append \id{suffix} \oplus C \inlineforall{C \in \id{propCmds}}$\;
\nl$\id{propCmds} \gets \emptyset$\;
\nl$las[\id{self}] \gets |v_a|$\;
\nl$\id{state} \gets (\atom{leader}, \atom{accept})$\;
\ForEach {$p \in \{p\in\Pi_0 \mid \id{lds}[p] \neq \bot\}$}{
\nl$\send{\msg{AcceptSync}{n_L, \atom{suffix}(v_a, \id{lds}[p]), \id{lds}[p]}}{p}$\;
}
}
}
\nl\Hdl{$\msg{Promise}{n, n', \id{suffix}_a, ld_a} \from{a} \st{n=n_L \wedge \id{state} = (\atom{leader}, \atom{accept})}$}
{
\nl$\id{lds}[a] \gets ld_a$\;
\nl$\send{\msg{AcceptSync}{n_L, \atom{suffix}(v_a, \id{lds}[a]), \id{lds}[a]}}{p}$\;
\nl\If{$l_c \neq 0$}{
\nl$\send{\msg{Decide}{l_c, n_L}}{a}$\tcc*{also inform what got decided already}
}
}
\end{algorithm}
\addtocounter{algocf}{-1}
\begin{algorithm}[htp]
\caption{Sequence Paxos Fail-Recovery -- Part 2}
\label{alg:seqpaxosrecovery2}
\tcc{Proposer Code (continued)}
\nl\Hdl{$\msg{Accepted}{n, l_a} \from{a} \st{n=n_L \wedge \id{state} = (\atom{leader}, \atom{accept})}$}
{
\nl$\id{las}[a] \gets l_a$\;
\nl$M \gets \{p \in \Pi \mid \id{las}[p] \neq \bot \wedge \id{las}[p] \geq l_a\}$\tcc*{support set for $l_a$}
\nl \If{$l_a > l_c \wedge |M| \geq \majority{\Pi}$}{
\nl $l_c \gets l_a$\;
\ForEach {$p \in \{p\in\Pi_0 \mid \id{lds}[p] \neq \bot\}$}{
\nl$\send{\msg{Decide}{l_c, n_L}}{p}$\tcc*{send length of chosen sequence}
}
}
}
\hrulefill \\
\tcc{Acceptor Code}
\nl\Hdl{$\msg{Prepare}{n, \id{ld}, \id{na}_L} \from{p}$}
{
\nl \If{$n_{prom} < n$}{
\nl$n_{prom} \gets n$\;
\nl$\id{state} \gets (\atom{follower}, \atom{prepare})$\;
\nl$\id{suffix} \gets \inlineif{n_a \geq \id{na}_L}{\atom{suffix}(v_a, \id{ld})}{\emptyseq}$\;
\nl$\send{\msg{Promise}{n, n_a, \id{suffix}, l_d}}{p}$\;
}
}
\nl \Hdl{$\msg{AcceptSync}{n, \id{suffix}, \id{ld}} \from{p} \st{\id{state} = (\atom{follower}, \atom{prepare})}$}
{
\nl \If{$n_{prom} = n$}{
\nl$n_a \gets n$\;
\nl$v_a \gets \atom{prefix}(v_a, \id{ld}) \append \id{suffix}$\;
\nl$\id{state} \gets (\atom{follower}, \atom{accept})$\;
\nl$\send{\msg{Accepted}{n, |v_a|}}{p}$\;
}
}
\nl \Hdl{$\msg{Accept}{n, C} \from{p} \st{\id{state} = (\atom{follower}, \atom{accept})}$}
{
\nl \If{$n_{prom} = n$}{
\nl$v_a \gets v_a \oplus C$\;
\nl$\send{\msg{Accepted}{n, |v_a|}}{p}$\;
}
}
\hrulefill \\
\tcc{Learner Code}
\nl\Hdl{$\msg{Decide}{l, n} \st{n=n_{prom}}$}
{
\nl \While{$l_d < l$}{
\nl$\trigger{\msg{Decide}{v_a[l_d]}}$\tcc*{assuming 0-based indexing}
\nl$l_d \gets l_d + 1$\;
}
}
\end{algorithm}
\clearpage

\subsection{Link Session Drop}
It was alluded to before, that TCP can be used to implement the FIFO Perfect Link abstraction, but only during a single session. This naturally begs the question of how to behave when a session drop event does occur. The semantics of session drop are equivalent to the \emph{omission} failures that occur during crash-recovery, that is an arbitrary suffix of the most recent messages will be lost. As this is clearly a sub-variant of a full crash-recovery event, i.e. a recovery without \emph{amnesia}, we will treat it very similarly. That is, at every process $p$ we will behave as follows in response to a $\msg{ConnectionLost}{q}$ event, indicating that the link session with process $q$ was dropped:
\setcounter{case}{0}
\begin{case}[$\id{state}=(\atom{leader}, \_)$] If we are leader, we simply continue as normal. The algorithm already handles the case where we are lacking a majority to proceed, so there is nothing else we can do except hope that $q$ comes back up and re-establishes connection in the future.
\end{case}
\begin{case}[$\id{state}=(\atom{follower}, \_) \wedge q=L$] If we are a follower and we lose connection to the leader, we are in the same situation as if we had recovered from a failure. Thus we will behave in the same way, moving to $\id{state}=(\atom{follower}, \atom{recover})$ and waiting for the $\msg{Leader}{L, n}$ message. Then behave just as in section \ref{ssec:failrec}.
\end{case}

The complete pseudocode for this version can be seen in algorithm \ref{alg:seqpaxossessionstate}.

\subsubsection{Correctness} The argument for correctness is the same as in section \ref{ssec:failrec}, considering that session loss in a sub-case of fail-recovery.

\begin{algorithm}[hbtp]
\caption{Sequence Paxos Fail-Recovery\&Session Loss -- State}
\label{alg:seqpaxossessionstate}
\textbf{Implements:} Sequence Consensus\\
\textbf{Requires:} FIFO Perfect Link, BLE \\
\textbf{Algorithm:} \\
\nl$\Pi$ \tcc*{set of processes}
\nl$\Pi_o \gets \Pi - \{\id{self}\}$\;
\nl$\id{state} \gets (\atom{follower}, \bot)$\tcc*{role and phase state}
\tcc{Proposer State}
\nl$n_L \gets 0$ \tcc*{leader's round number}
\nl$\id{promises}$ $\gets \emptyset$\;
\nl$\id{las} \gets [0]^{|\Pi|}$\tcc*{length of longest accepted sequence per acceptor}
\nl$\id{lds} \gets [\bot]^{|\Pi|}$\tcc*{length of longest known decided sequence per acceptor}
\nl$\id{propCmds} \gets \emptyset$\tcc*{set of commands that need to be appended to the log}
\nl$l_c \gets 0$ \tcc*{length of longest chosen sequence}
\tcc{Acceptor State}
\nl$\persist{n_{prom}}\gets 0$ \tcc*{promise not to accept in lower rounds}
\nl$\persist{(n_a, v_a)} \gets (0, \emptyseq)$ \tcc*{round number and sequence accepted}
\tcc{Learner State}
\nl$\persist{l_d} \gets 0$ \tcc*{length of the decided sequence}
\end{algorithm}
\addtocounter{algocf}{-1}
\begin{algorithm}[htp]
\caption{Sequence Paxos Fail-Recovery\&Session Loss -- Part 1}
\label{alg:seqpaxossession1}
\tcc{General Code}
\nl\Hdl{$\msg{Leader}{L, n}$}
{
\eIf{$\id{self} = L \wedge n > n_L \wedge n > n_{prom}$}{
\nl$(n_L, n_{prom}) \gets (n, n)$\;
\nl$\id{promises}$ $\gets \{ (\id{self}, n_a, \atom{suffix}(v_a, l_d)) \}$\;
\nl$\id{las} \gets [0]^{|\Pi|}$\;
\nl$\id{lds} \gets [\bot]^{|\Pi|}; \id{lds}[\id{self}] \gets l_d$\;
\nl$l_c \gets 0$\;
\nl$\id{state} \gets (\atom{leader}, \atom{prepare})$\;
\ForEach {$p \in \Pi_o$}{
\nl$\send{\msg{Prepare}{n_L, l_d, n_a}}{p}$\;
}
}{
\eIf{$\id{state}=(\_, \atom{recover})$}{
\nl$\send{\msgatom{PrepareReq}}{L}$\;
}{
\nl$\id{state} \gets (\atom{follower}, \id{state}.2)$\;
}
}
}

\HiLi\nl\Hdl{$\msg{ConnectionLost}{q} \st{\id{state} = (\atom{follower}, \_) \wedge q=L}$}
{
\HiLi\nl$\id{state} \gets (\atom{follower}, \atom{recover})$\;
}
\tcc{Leader Code}
\nl\Hdl{$\msgatom{PrepareReq} \from{a} \st{\id{state} = (\atom{leader}, \_)}$}
{
\nl$\send{\msg{Prepare}{n_L, l_d, n_a}}{a}$\;
}

\nl\Hdl{$\msg{Propose}{C} \st{\id{state} = (\atom{leader}, \atom{prepare})}$}
{
\nl$\id{propCmds} \gets \id{propCmds \cup \{C\}}$\;
}
\nl\Hdl{$\msg{Propose}{C} \st{\id{state} = (\atom{leader}, \atom{accept})}$}
{
\nl$v_a \gets v_a \oplus C$\;
\nl$las[\id{self}] \gets |v_a|$\;
\ForEach {$p \in \{p\in\Pi_0 \mid \id{lds}[p] \neq \bot\}$}{
\nl$\send{\msg{Accept}{n_L, C}}{p}$\;
}
}

\nl\Hdl{$\msg{Promise}{n, n', \id{suffix}_a, ld_a} \from{a} \st{n=n_L \wedge \id{state} = (\atom{leader}, \atom{prepare})}$}
{
\nl$\id{promises} \gets \id{promises} \cup \{(a, n', \id{suffix}_a)\}$\;
\nl$\id{lds}[a] \gets ld_a$\;
\nl \If{$|\id{promises}| = \majority{\Pi}$}{
\nl$\id{suffix} \gets \atom{maxValue}(\id{promises})$\tcc*{suffix with max $n$, longest if equal}
\tcc{adopt $v_d \append \id{suffix}$ and append commands}
\nl$v_a \gets \atom{prefix}(v_a, l_d) \append \id{suffix} \oplus C \inlineforall{C \in \id{propCmds}}$\;
\nl$\id{propCmds} \gets \emptyset$\;
\nl$las[\id{self}] \gets |v_a|$\;
\nl$\id{state} \gets (\atom{leader}, \atom{accept})$\;
\ForEach {$p \in \{p\in\Pi_0 \mid \id{lds}[p] \neq \bot\}$}{
\nl$\send{\msg{AcceptSync}{n_L, \atom{suffix}(v_a, \id{lds}[p]), \id{lds}[p]}}{p}$\;
}
}
}
\nl\Hdl{$\msg{Promise}{n, n', \id{suffix}_a, ld_a} \from{a} \st{n=n_L \wedge \id{state} = (\atom{leader}, \atom{accept})}$}
{
\nl$\id{lds}[a] \gets ld_a$\;
\nl$\send{\msg{AcceptSync}{n_L, \atom{suffix}(v_a, \id{lds}[a]), \id{lds}[a]}}{p}$\;
\nl\If{$l_c \neq 0$}{
\nl$\send{\msg{Decide}{l_c, n_L}}{a}$\tcc*{also inform what got decided already}
}
}
\end{algorithm}
\addtocounter{algocf}{-1}
\begin{algorithm}[htp]
\caption{Sequence Paxos Fail-Recovery\&Session Loss -- Part 2}
\label{alg:seqpaxossession2}
\tcc{Leader Code (continued)}
\nl\Hdl{$\msg{Accepted}{n, l_a} \from{a} \st{n=n_L \wedge \id{state} = (\atom{leader}, \atom{accept})}$}
{
\nl$\id{las}[a] \gets l_a$\;
\nl$M \gets \{p \in \Pi \mid \id{las}[p] \neq \bot \wedge \id{las}[p] \geq l_a\}$\tcc*{support set for $l_a$}
\nl \If{$l_a > l_c \wedge |M| \geq \majority{\Pi}$}{
\nl $l_c \gets l_a$\;
\ForEach {$p \in \{p\in\Pi_0 \mid \id{lds}[p] \neq \bot\}$}{
\nl$\send{\msg{Decide}{l_c, n_L}}{p}$\tcc*{send length of chosen sequence}
}
}
}
\hrulefill \\
\tcc{Acceptor Code}
\nl\Hdl{$\msg{Prepare}{n, \id{ld}, \id{na}_L} \from{p}$}
{
\nl \If{$n_{prom} < n$}{
\nl$n_{prom} \gets n$\;
\nl$\id{state} \gets (\atom{follower}, \atom{prepare})$\;
\nl$\id{suffix} \gets \inlineif{n_a \geq \id{na}_L}{\atom{suffix}(v_a, \id{ld})}{\emptyseq}$\;
\nl$\send{\msg{Promise}{n, n_a, \id{suffix}, l_d}}{p}$\;
}
}
\nl \Hdl{$\msg{AcceptSync}{n, \id{suffix}, \id{ld}} \from{p} \st{\id{state} = (\atom{follower}, \atom{prepare})}$}
{
\nl \If{$n_{prom} = n$}{
\nl$n_a \gets n$\;
\nl$v_a \gets \atom{prefix}(v_a, \id{ld}) \append \id{suffix}$\;
\nl$\id{state} \gets (\atom{follower}, \atom{accept})$\;
\nl$\send{\msg{Accepted}{n, |v_a|}}{p}$\;
}
}
\nl \Hdl{$\msg{Accept}{n, C} \from{p} \st{\id{state} = (\atom{follower}, \atom{accept})}$}
{
\nl \If{$n_{prom} = n$}{
\nl$v_a \gets v_a \oplus C$\;
\nl$\send{\msg{Accepted}{n, |v_a|}}{p}$\;
}
}
\hrulefill \\
\tcc{Learner Code}
\nl\Hdl{$\msg{Decide}{l, n} \st{n=n_{prom}}$}
{
\nl \While{$l_d < l$}{
\nl$\trigger{\msg{Decide}{v_a[l_d]}}$\tcc*{assuming 0-based indexing}
\nl$l_d \gets l_d + 1$\;
}
}
\end{algorithm}
\clearpage

\section{Reconfiguration}
\label{sec:reconf}
Having dealt with transient failures, we must now deal with permanent node failures. This can be physical hardware failing to the point that a node has to be permanently removed, but it could also simply be the desire to add or remove nodes to a running system in order to deal with changed load. As \emph{reconfiguration} has many use cases in a practical system, we will separate out the \emph{policy}, that is ``why'' and ``when'' we are reconfiguring, from the \emph{mechanism}, i.e. ``how'' we are reconfiguring, given that the decision has been made already. Policy questions a very application dependent and could be anything from a human operator making decisions in a control room, to a fully automated system making decisions based on some monitoring information. In this section, we will only cover the mechanisms needed to reconfigure a \emph{Sequence Paxos} group.

\subsection{Configurations}
We will call a ``group membership instance'' a \emph{configuration} $c_i$. For example, for four processes $p_1, \ldots, p_4$ we might start in configuration $c_0 = \{p_1, p_2, p_3\}$, but at some point the policy decides to move to a new configuration $c_1 = \{p_1, p_2, p_4\}$. We make no restrictions on the membership in each configuration, that is, in general it might happen that $c_0 \cap c_1 = \emptyset$. Most of the time, though, we will simply replace a single node that is considered failed by the policy.\\
We model our system such that every configuration $c_i$ is a logically separate instance of \emph{Sequence Paxos}, with its own instance of \emph{Ballot Leader Election}. Each process $p_j\in c_i$ acts as a replica $r_{ij}$ in $c_i$. Thus, a process $p_j$ may be a replica in multiple configurations at the same time. Reusing the same example as before, we would have configuration $c_0 = \{r_{01}, r_{02}, r_{03}\}$ and later $c_1 = \{r_{11}, r_{12}, r_{14}\}$. In this case, for example, $p_1$ is part of two configurations in $\{r_{01},r_{11}\}$.

The RSM executes in a configuration until a reconfiguration event occurs, then it moves to the new configuration. At each process this transitions happens asynchronously, but only a single configuration is (locally) \emph{active} (or \emph{running}) at a time, that is it can \emph{extend} its sequence. A new configuration is considered (globally) \emph{active}, once it has a majority of active members.

In order to mark the end of a configuration $c_i$, we will issue a special command called the \emph{stop-sign} $SS_i$. Once a leader $p_L$ proposes a sequence  $\sigma_i$ containing $SS_i$ in configuration $c_i$, it must be such that $SS_i$ is the last command in $\sigma_i$ and $p_L$ may not issue a longer sequence in $c_i$. Once $\sigma_i$ is \emph{decided} no proposer may ever extend it, thus we call $\sigma_i$ the \emph{final sequence} of $c_i$ and we call $c_i$ \emph{stopped}.\\
When the $\sigma_i$ is decided in $c_i$ by at least one process, the new configuration $c_{i+1}$ can start, as it is guaranteed that $\sigma_i$ will not change. The stop sign $SS_i$ for configuration $c_i$ contains the all the information necessary to start $c_{i+1}$. Concretely that is $\Pi_{i+1}$, the set of processes in $c_{i+1}$, $i$, the \emph{configuration number}, and for each process $p_j\in\Pi_{i+1}$ its \emph{replica identifier} $r_{(i+1) j}$. We want to use $\sigma_i$ as initial sequence for all replicas in $c_{i+1}$. There are three cases that we have to deal with at every process $p_j\in\Pi_{i+1}$:
\setcounter{case}{0}
\begin{case}[$c_i = c_0$] We are starting the first configuration and there is no previous sequence. The we pick $\emptyseq$ as the initial sequence on all replicas. 
\end{case}
\begin{case}[$c_i \neq c_0 \wedge p_j\in\Pi_{i}$] This process is a replica in both the $c_i$ and $c_{I+1}$. Once it has locally decided $\sigma_i$ in the instance for configuration $c_i$, it will pass $\sigma_i$ as an initial parameter to the instance for configuration $c_{i+1}$ locally. (Since $\sigma_i$ is immutable, it could even pass a reference, thus sharing the memory in an implementation that allows such things.)
\end{case}
\begin{case}[$c_i \neq c_0 \wedge p_j\notin\Pi_{i}$] This is a process that does not have the initial sequence locally already. It must fetch it from some other node, either from the old configuration or from a shared persistent storage, before it can start up in the new configuration. This transfer process can be very time-consuming if the log is large. Practically, it is advisable to begin this process in parallel to the old configuration still running and only once the new nodes are close to being caught up, issue the stop-sign command. Additional optimisations such as state compression (snapshotting) and garbage collection on the log and state machine's state are highly recommended (s. section \ref{sec:gc}).
\end{case}

To make the proposed changes work with our \emph{Sequence Paxos} algorithm we will additionally extend the \emph{round numbers} to contain the \emph{configuration number} as well, such that rounds in higher configurations will always be ordered higher than rounds from older configurations. Thus instead having round number $n = b$ for a ballot number $b$, we will now have $n = (i, b)$ for a configuration $c_i$.\\

Algorithm \ref{alg:seqpaxosreconfstate} describes the execution of a replica after it has acquire its initial sequences $\sigma_{i-1}$. The startup stage is assumed to be handled by an external component, which starts the replica after obtaining the initial sequence. Given the many and application-dependent options for handling that part of the algorithm, we chose not provide pseudocode for it here.

\subsubsection{Correctness} Since configurations are totally ordered, our new round numbers that include the configration identifier are also totally ordered across configurations. Furthermore, the acceptor state at the start of a new configuration is the same as the state of the acceptors at the end of the last configuration. In this way we maintain the invariant, that if a sequence $v$ is issued in round $n=(i, b)$, then $v$ is an extension of all sequences chosen in previous rounds $n' \leq n$.

\subsubsection{Performance} The performance of the proposed approach depends mostly on how efficient the distribution of the final sequences to the members of the next configuration is. During an active configuration the described method has little or no performance impact at all. However, the necessity to keep replicas in old configurations around to catch up recovering members introduces a certain amount of clutter over time, which we will have to deal with at some point.

\begin{algorithm}[hbtp]
\caption{Sequence Paxos Reconfiguration -- State\&General}
\label{alg:seqpaxosreconfstate}
\textbf{Implements:} Sequence Consensus\\
\textbf{Requires:} FIFO Perfect Link, BLE \\
\textbf{Algorithm:} \\
\HiLi\nl$c_i$ \tcc*{configuration this replica is running in}
\HiLi\nl$\Pi_i$ \tcc*{set of processes in configuration $c_i$}
\HiLi\nl$R \gets \{r_{ij} \mid p_j\in\Pi_i\}$ \tcc*{set of replicas in configuration $c_i$}
\HiLi\nl$R_o \gets R - \{\id{self}\}$\;
\HiLi\nl$\id{rself} \gets r_{ij} \st{\id{self}=p_j\in\Pi_i}$\tcc*{our own replica id for this configuration}
\HiLi\nl$\sigma_{i-1}$ \tcc*{the final sequence from the previous configuration or $\emptyseq$ if $i=0$}
\nl$\id{state} \gets (\atom{follower}, \bot)$\tcc*{role and phase state}
\tcc{Proposer State}
\HiLi\nl$n_L \gets (i, 0)$ \tcc*{leader's round number}
\nl$\id{promises}$ $\gets \emptyset$\;
\HiLi\nl$\id{las} \gets [ \ |\sigma_{i-1}| \ ]^{|R|}$\tcc*{length of longest accepted sequence per acceptor}
\HiLi\nl$\id{lds} \gets [\bot]^{|R|}$\tcc*{length of longest known decided sequence per acceptor}
\nl$\id{propCmds} \gets \emptyset$\tcc*{set of commands that need to be appended to the log}
\HiLi\nl$l_c \gets |\sigma_{i-1}|$ \tcc*{length of longest chosen sequence}
\tcc{Acceptor State}
\HiLi\nl$\persist{n_{prom}}\gets (i, 0)$ \tcc*{promise not to accept in lower rounds}
\HiLi\nl$\persist{(n_a, v_a)} \gets ((i,0), \sigma_{i-1})$ \tcc*{round number and sequence accepted}
\tcc{Learner State}
\HiLi\nl$\persist{l_d} \gets |\sigma_{i-1}|$ \tcc*{length of the decided sequence}
\hrulefill \\
\tcc{General Code}
\nl\Fn{$\atom{stopped}()$}{$\Return \ \atom{last}(v_a) = SS_i$\;}
\nl\Hdl{$\msg{Leader}{L, b}$}
{
\HiLi\nl$n \gets (i, b)$\;
\eIf{$\id{self} = L \wedge n > n_L \wedge n > n_{prom}$}{
\nl$(n_L, n_{prom}) \gets (n, n)$\;
\HiLi\nl$\id{promises}$ $\gets \{ (\id{rself}, n_a, \atom{suffix}(v_a, l_d)) \}$\;
\HiLi\nl$\id{las} \gets [|\sigma_{i-1}|]^{|R|}$\;
\HiLi\nl$\id{lds} \gets [\bot]^{|R|}; \id{lds}[\id{rself}] \gets l_d$\;
\HiLi\nl$l_c \gets |\sigma_{i-1}|$\;
\nl$\id{state} \gets (\atom{leader}, \atom{prepare})$\;
\HiLi\ForEach {$r \in R_o$}{
\HiLi\nl$\send{\msg{Prepare}{n_L, l_d, n_a}}{r}$\;
}
}{
\eIf{$\id{state}=(\_, \atom{recover})$}{
\nl$\send{\msgatom{PrepareReq}}{L}$\;
}{
\nl$\id{state} \gets (\atom{follower}, \id{state}.2)$\;
}
}
}

\nl\Hdl{$\msg{ConnectionLost}{q} \st{\id{state} = (\atom{follower}, \_) \wedge q=L}$}
{
\nl$\id{state} \gets (\atom{follower}, \atom{recover})$\;
}
\end{algorithm}
\addtocounter{algocf}{-1}
\begin{algorithm}[htp]
\caption{Sequence Paxos Reconfiguration -- Leader (1)}
\label{alg:seqpaxosreconf1}
\tcc{Leader Code}
\nl\Hdl{$\msgatom{PrepareReq} \from{a} \st{\id{state} = (\atom{leader}, \_)}$}
{
\nl$\send{\msg{Prepare}{n_L, l_d, n_a}}{a}$\;
}
\nl\Hdl{$\msg{Propose}{C} \st{\id{state} = (\atom{leader}, \atom{prepare})}$}
{
\nl$\id{propCmds} \gets \id{propCmds \cup \{C\}}$\;
}
\HiLi\nl\Hdl{$\msg{Propose}{C} \st{\id{state} = (\atom{leader}, \atom{accept}) \wedge \neg\atom{stopped}()}$}
{
\nl$v_a \gets v_a \oplus C$\;
\HiLi\nl$las[\id{rself}] \gets |v_a|$\;
\HiLi\ForEach {$p \in \{r\in R_0 \mid \id{lds}[r] \neq \bot\}$}{
\HiLi\nl$\send{\msg{Accept}{n_L, C}}{r}$\;
}
}
\nl\Hdl{$\msg{Promise}{n, n', \id{suffix}_a, ld_a} \from{a} \st{n=n_L \wedge \id{state} = (\atom{leader}, \atom{prepare})}$}
{
\nl$\id{promises} \gets \id{promises} \cup \{(a, n', \id{suffix}_a)\}$\;
\nl$\id{lds}[a] \gets ld_a$\;
\nl \If{$|\id{promises}| = \majority{\Pi}$}{
\nl$\id{suffix} \gets \atom{maxValue}(\id{promises})$\tcc*{suffix with max $n$, longest if equal}
\tcc{adopt $v_d \append \id{suffix}$ and append commands}
\nl$v_a \gets \atom{prefix}(v_a, l_d) \append \id{suffix}$\;
\HiLi \eIf{$SS_i = \atom{last}(v_a)$}{
\HiLi\nl$\id{propCmds} \gets \emptyset$\tcc*{commands will never be decided}
}{
\HiLi \eIf{$SS_i \in \id{propCmds}$}{
\tcc{Could also just drop other outstanding commands instead of ordering them before $SS_i$}
\HiLi\nl$v_a \gets v_a \oplus C \inlineforall{C \in \id{propCmds} - \{SS_i\}}$\;
\HiLi\nl$v_a \gets v_a \oplus SS_i$\;
}{
\HiLi\nl$v_a \gets v_a \oplus C \inlineforall{C \in \id{propCmds}}$\;
}
}
\nl$las[\id{self}] \gets |v_a|$\;
\nl$\id{state} \gets (\atom{leader}, \atom{accept})$\;
\HiLi\ForEach {$r \in \{r\in R_0 \mid \id{lds}[r] \neq \bot\}$}{ 
\HiLi\nl$\send{\msg{AcceptSync}{n_L, \atom{suffix}(v_a, \id{lds}[r]), \id{lds}[r]}}{r}$\;
}
}
}
\end{algorithm}
\addtocounter{algocf}{-1}
\begin{algorithm}[htp]
\caption{Sequence Paxos Reconfiguration -- Leader (2)}
\label{alg:seqpaxosreconf2}
\tcc{Leader Code (continued)}
\nl\Hdl{$\msg{Promise}{n, n', \id{suffix}_a, ld_a} \from{a} \st{n=n_L \wedge \id{state} = (\atom{leader}, \atom{accept})}$}
{
\nl$\id{lds}[a] \gets ld_a$\;
\nl$\send{\msg{AcceptSync}{n_L, \atom{suffix}(v_a, \id{lds}[a]), \id{lds}[a]}}{a}$\;
\nl\If{$l_c \neq |\sigma_{i-1}|$}{
\nl$\send{\msg{Decide}{l_c, n_L}}{a}$\tcc*{also inform what got decided already}
}
}
\nl\Hdl{$\msg{Accepted}{n, l_a} \from{a} \st{n=n_L \wedge \id{state} = (\atom{leader}, \atom{accept})}$}
{
\nl$\id{las}[a] \gets l_a$\;
\nl$M \gets \{p \in \Pi \mid \id{las}[p] \neq \bot \wedge \id{las}[p] \geq l_a\}$\tcc*{support set for $l_a$}
\nl \If{$l_a > l_c \wedge |M| \geq \majority{\Pi}$}{
\nl $l_c \gets l_a$\;
\ForEach {$p \in \{p\in\Pi_0 \mid \id{lds}[p] \neq \bot\}$}{
\nl$\send{\msg{Decide}{l_c, n_L}}{p}$\tcc*{send length of chosen sequence}
}
}
}
\end{algorithm}
\addtocounter{algocf}{-1}
\begin{algorithm}[htp]
\caption{Sequence Paxos Reconfiguration -- Acceptor \& Learner}
\label{alg:seqpaxosreconf3}

\tcc{Acceptor Code}
\nl\Hdl{$\msg{Prepare}{n, \id{ld}, \id{na}_L} \from{p}$}
{
\nl \If{$n_{prom} < n$}{
\nl$n_{prom} \gets n$\;
\nl$\id{state} \gets (\atom{follower}, \atom{prepare})$\;
\nl$\id{suffix} \gets \inlineif{n_a \geq \id{na}_L}{\atom{suffix}(v_a, \id{ld})}{\emptyseq}$\;
\nl$\send{\msg{Promise}{n, n_a, \id{suffix}, l_d}}{p}$\;
}
}
\nl \Hdl{$\msg{AcceptSync}{n, \id{suffix}, \id{ld}} \from{p} \st{\id{state} = (\atom{follower}, \atom{prepare})}$}
{
\nl \If{$n_{prom} = n$}{
\nl$n_a \gets n$\;
\nl$v_a \gets \atom{prefix}(v_a, \id{ld}) \append \id{suffix}$\;
\nl$\id{state} \gets (\atom{follower}, \atom{accept})$\;
\nl$\send{\msg{Accepted}{n, |v_a|}}{p}$\;
}
}
\nl \Hdl{$\msg{Accept}{n, C} \from{p} \st{\id{state} = (\atom{follower}, \atom{accept})}$}
{
\nl \If{$n_{prom} = n$}{
\nl$v_a \gets v_a \oplus C$\;
\nl$\send{\msg{Accepted}{n, |v_a|}}{p}$\;
}
}
\hrulefill \\
\tcc{Learner Code}
\nl\Hdl{$\msg{Decide}{l, n} \st{n=n_{prom}}$}
{
\nl \While{$l_d < l$}{
\nl$C \gets v_a[l_d]$\tcc*{assuming 0-based indexing}
\nl$\trigger{\msg{Decide}{C}}$\;
\nl$l_d \gets l_d + 1$\;
}
}
\end{algorithm}

\clearpage

\section{Garbage Collection}
\label{sec:gc}
As was alluded to in the last paragraph, we have been silently ignoring a certain build-up of state over the lifetime of systems running the algorithms presented so far. In particular there are two areas, the log $v_a$ and the number of stopped configurations $c_i$, that continue to grow without bounds throughout the system's lifetime. In this section we will explore some ideas on how to manage these issues in a real deployment.

\subsection{Snapshots and Truncating the Log}
In the most general case, we can not truncate log, as we may have to catch up a new replica that was added in a new configuration. That is, without any knowledge about how the state machine $S$ on top of the replicated log works, the only way to catch up a new member is to replay the whole log $v_d$ onto $S$' initial state $s_0$. However, if $S$ allows us to persist intermediate states $s_i$, which we shall call \emph{snapshots}, we may in fact truncate the log up to all the commands included in $s_i$, an then transfer $s_i$ and the remaining log to a new replicate to catch it up.

For the vast majority of RSMs this approach will be orders of magnitude faster than transferring and replaying the whole log. Consider for example a key-value store on top of a replicated log, such that the commands are $\atom{put}$ and $\atom{get}$. The log might contain millions of messages after minutes already, but the state of the store would not include any $\atom{get}$ commands and it would only grow in the size of $\atom{put}$ commands on separate keys (or the size of the values, if that is not constant). It is easy to see that for most workloads (which tend to be $\atom{get}$-heavy) the size of any state $s_i$ is going to be orders of magnitude smaller than the size of the decided sequence $v_d$.

We must however ensure, that every replica has persisted its snapshot before we truncate the log anywhere to make sure that we can still catch up replicas that crashed and are recovering during the period between snapshot and truncation\footnote{This isn't technically true, we could take a majority snapshot and if a node recovers too late we could change configuration without changing membership and re-use the mechanism for distributing the initial sequence to catch it up.}.
To that end we leave the frequency of snapshots up to the implementation of $S$, but require the interface to inform the \emph{Sequence Paxos} algorithm when a snapshot is finished locally at $p_j$ by proposing a special $\msg{Snapshot}{j, k, l_k}$ command where $k$ is the identifier of the snapshot $\dot{s}_k$ and $l_k$ a pointer to the last command in $v_d$ that is included in the state $s_{l_k}$ being captured by $\dot{s}_k$. In general we can truncate $v_a$ up to any position $m$ where we have decided $\msg{Snapshot}{j, k, l_k}$ with $l_k \geq m$ for all $p_j\in\Pi_i$. However, it is most convenient to deterministically take snapshots at the same point in the log at all replicas, such that we only need to keep track of the largest $l_k$ seen from all replicas. Once the log is truncated, we need to take care to translate offset-based pointers like $l_d$ by the last $l_k$ before doing lookups into $v_a$, such that that $v_a[l_d]$ before truncation becomes $v_a[l_d - l_k]$. Similarly all usages of $|v_a|$ need to be translated to $|v_a|+l_k$.

Additionally, to avoid integer overflows of $l_d$, for example, we can reset all counters to $0$ after reconfiguration, iff we make sure the final sequence is always transferred as a pure snapshot (and not a combination of snapshot and truncated log). 

One issue we have to deal with during log truncation has to do with command duplication as we described in section \ref{ssec:seqpax1}. Recall that we defined the append $\oplus$ implementation \emph{without duplicates}, such that it would check the log before appending a command $C$, to see if i$C$ already existed and in such a case skip it. If we are truncating the log, we can not check it for commands anymore, and depending on the RSM implementation it may in fact be impossible to see from a snapshot whether or not a command $C$ was already applied to it. For example, in a key-value store, if the snapshot is the mapping from keys to current values, then even a $\atom{PUT}(k, v)$ may have been overwritten in the snapshot already, just making its prior application impossible to deduce. For this reason we must find another way to deduplicate commands, if we are to truncate the log.\\
If we assume that all clients execute \emph{sequentially}, that is sending only a single command $C$ at a time and waiting for it be acknowledged (possibly resending $C$ occasionally after timeouts) before sending a new command $C'$, we can use the following mechanism for deduplication: The RSM on each server maintains a mapping from a client id $p_c$ to the last command $C_c$ that was submitted by $p_c$ and decided, together with the result of executing $C_c$ on the RSM state $S$ (if it is an operation with a result, such as a $\atom{GET}$). Whenever a new command $C'_c$ is from $p_c$ is decided, we check if $C_c = C'_c$ and if so we do not execute it again, but simply send the stored result to $p_c$. If, on the other hand, $C_c \neq C'_c$ then we execute it on $S$ and store it and the result in the map, replacing $C_c$. In this way we never execute duplicate commands, as for sequential clients duplicate commands must directly follow each other without a different command occurring in between. It is assumed that clients simply ignore additional responses to commands they consider complete.\\
When doing snapshots now, we must ensure that during log replay after reconfiguration (or recovery) we also do not apply duplicate commands. To achieve this we must store the state of this client map at the time a snapshot is taken together with snapshot. Then we start from the snapshot we also load its client map and apply commands replayed from the log, as if they were newly incoming commands, that is checking for duplicates and only applying the first instance of each command together with its result.

\subsection{Configuration Cleanup}
Once a new configuration $c_i$ is started, the previous configuration $c_{i-1}$ is not needed anymore. It seems trivial to just shut down all resources related to it, but the possibility of long-running network partitions makes cleanup of these replicas a surprisingly difficult problem without external input. To see why this is the case, consider the following scenario: In configuration $c_1$ we have three replicas $\Pi_1=\{p_1, p_2, p_3\}$ and after running for a while $p_3$ is disconnected from the rest. The reconfiguration policy $\rho$ decides that $p_3$ is to be replaced by a new process $p_4$ by transitioning to configuration $c_2$ with $\Pi_2 = \{p_1, p_2, p_4\}$. Eventually $SS_1$ is decided in $c_1$ and the final sequence $\sigma_1$ is transferred to $p_4$ (and locally  to $p_1$ and $p_2$) allowing the replicas in $c_2$ to start up. Now fast forward a few failures and maybe we are in configuration $c_5$ with $\Pi_5=\{p_4, p_5, p_6\}$ and we have shut down (either on purpose or due to failure) the replicas at $p_1$ and $p_2$. However, due to the partition we have been unable to inform $p_3$ so far that it is not needed anymore. Say at this point suddenly $p_3$ reconnects. Being alone it can't do anything wrong, but it has no idea that its supposed to be shutting down. There is no one left it can talk to, $p_1$ and $p_2$ being dead, but it can't just shutdown either, since it might just still be disconnected from the other two. Without some way to get external input $p_3$ is going to stay in this orphaned state forever.

There are multiple way to deal with this, but in the end they all come down to outsourcing the shutdown decision to some kind of policy $\rho$. If $\rho$ happens to be a human administrator, they can simply physically go to the disconnected machine $p_3$ and shut down the process for the orphaned replica. Obviously, this isn't very convenient. A more automated system could make use of infrastructure existing at the data-centre for service discovery (e.g., ARP) or name resolution (e.g., DNS) to allow $p_3$ to discover another working process such as $p_4$ to acquire information about what happened to its configuration by inspecting the active log with a client API, for example.

\clearpage
\pagebreak

\section{Literature}
\label{sec:lit}
While Paxos was originally presented by Leslie Lamport in \cite{lamport1998part}, the way we describe it in these lecture notes comes from a later paper~\cite{lamport2001paxos}, with some additional ideas borrowed from \cite{de1997revisiting}. The description of sequence consensus, which was originally presented by Lamport in \cite{lamport2005generalized}, is purposefully aligned with the way that \emph{Raft}~\cite{ongaro2014search} is presented, to make it easy to compare the two implementations. Additionally, the ideas for the Sequence Paxos reconfiguration are based on ``Stoppable Paxos''~\cite{lamport2010reconfiguring}.

\section{Conclusion}
\label{sec:conc}
We have described how to get from single value Paxos via fail-stop Sequence Paxos, fail-recovery Sequence Paxos, to a reconfigurable implementation of the Sequence Consensus abstraction that works in the fail-recovery model. We have also discussed systems issues like garbage collection, snapshots and log truncations, and state transfers. The final algorithm we presented is ready to be be used as is, or adapted to a particular RSM implementation for efficiency.

\bibliography{references}

\begin{thebibliography}{10}
\providecommand{\url}[1]{\texttt{#1}}
\providecommand{\urlprefix}{URL }
\providecommand{\doi}[1]{https://doi.org/#1}

\bibitem{burrows2006chubby}
Burrows, M.: The chubby lock service for loosely-coupled distributed systems.
  In: Proceedings of the 7th symposium on Operating systems design and
  implementation. pp. 335--350. USENIX Association (2006)

\bibitem{cachin2011introduction}
Cachin, C., Guerraoui, R., Rodrigues, L.: Introduction to reliable and secure
  distributed programming. Springer Science \& Business Media (2011)

\bibitem{de1997revisiting}
De~Prisco, R., Lampson, B., Lynch, N.: Revisiting the paxos algorithm. In:
  International Workshop on Distributed Algorithms. pp. 111--125. Springer
  (1997)

\bibitem{hunt2010zookeeper}
Hunt, P., Konar, M., Junqueira, F.P., Reed, B.: Zookeeper: Wait-free
  coordination for internet-scale systems. In: USENIX annual technical
  conference. vol.~8. Boston, MA, USA (2010)

\bibitem{kroll2013load}
Kroll, L.: Load balancing in a distributed storage system for big and small
  data (2013)

\bibitem{lamport2005generalized}
Lamport, L.: Generalized consensus and paxos  (2005)

\bibitem{lamport2010reconfiguring}
Lamport, L., Malkhi, D., Zhou, L.: Reconfiguring a state machine. SIGACT News
  \textbf{41}(1),  63--73 (2010)

\bibitem{lamport1998part}
Lamport, L., et~al.: The part-time parliament. ACM Transactions on Computer
  systems  \textbf{16}(2),  133--169 (1998)

\bibitem{lamport2001paxos}
Lamport, L., et~al.: Paxos made simple. ACM Sigact News  \textbf{32}(4),
  18--25 (2001)

\bibitem{ongaro2014search}
Ongaro, D., Ousterhout, J.: In search of an understandable consensus algorithm.
  In: 2014 $\{$USENIX$\}$ Annual Technical Conference
  ($\{$USENIX$\}$$\{$ATC$\}$ 14). pp. 305--319 (2014)

\end{thebibliography}
\end{document}